\documentclass[preprint]{aastex} 

\shorttitle{clusters}
\shortauthors{Adams, Proszkow, Fatuzzo, and Myers}

\newcommand{\be}{\begin{equation}}
\newcommand{\ee}{\end{equation}}
  
\newcommand{\fone}{ {\cal F}_1 } 
\newcommand{\lexp}{ \langle L_{FUV} \rangle_\ast } 
\newcommand{\sigbar}{ \langle \sigma \rangle_{\rm fuv} } 
\newcommand{\cross}{ {\langle \sigma \rangle}} 
\newcommand{\rstar}{ {R_{c \ast} } } 
\def\lta{\,\raise 0.3 ex\hbox{$ < $}\kern -0.75 em
 \lower 0.7 ex\hbox{$\sim$}\,}
\def\gta{\,\raise 0.3 ex\hbox{$ > $}\kern -0.75 em
 \lower 0.7 ex\hbox{$\sim$}\,} 

\begin{document}

\title{Early Evolution of Stellar Groups and Clusters: \\
Environmental Effects on Forming Planetary Systems} 

\author{Fred C. Adams,$^{1,2}$ Eva M. Proszkow,$^{1}$ 
Marco Fatuzzo,$^3$  and Philip C. Myers$^4$} 
 
\affil{$^1$Michigan Center for Theoretical Physics, University of Michigan, Ann Arbor, MI 48109}  

\affil{$^2$Astronomy Department, University of Michigan, Ann Arbor, MI 48109}

\affil{$^3$Physics Department, Xavier University, Cincinnati, OH 45207} 

\affil{$^4$Harvard Smithsonian Center for Astrophysics, 60 Garden Street, 
Cambridge, MA 02138} 

\email{fca@umich.edu} 

\begin{abstract} 

This paper studies the dynamical evolution of young groups/clusters,
with $N = 100 - 1000$ members, from their embedded stage out to ages
of $\sim 10$ Myr.  We use $N$-body simulations to explore how their
evolution depends on the system size $N$ and the initial conditions.
Motivated by recent observations suggesting that stellar groups begin
their evolution with subvirial speeds, this study compares subvirial
starting states with virial starting states. Multiple realizations of
equivalent cases (100 simulations per initial condition) are used to
build up a robust statistical description of these systems, e.g., the
probability distribution of closest approaches, the mass profiles, and
the probability distribution for the radial location of cluster
members.  These results provide a framework from which to assess the
effects of groups/clusters on the processes of star and planet
formation, and to study cluster evolution. The distributions of radial
positions are used in conjunction with the probability distributions
of the expected FUV luminosities (calculated here as a function of
cluster size $N$) to determine the radiation exposure of circumstellar
disks. The distributions of closest approaches are used in conjunction
with scattering cross sections (calculated here as a function of
stellar mass using $\sim10^5$ Monte Carlo scattering experiments) to
determine the probability of disruption for newly formed solar
systems.  We use the nearby cluster NGC 1333 as a test case in this
investigation. The main conclusion of this study is that clusters in
this size range have only a modest effect on forming planetary
systems.  The interaction rates are low so that the typical solar
system experiences a single encounter with closest approach distance
$b \sim 1000$ AU. The radiation exposure is also low, with median FUV
flux $G_0 \sim 900$ (1.4 erg s$^{-1}$ cm$^{-2}$), so that
photoevaporation of circumstellar disks is only important beyond 30
AU. Given the low interaction rates and modest radiation levels, we
suggest that solar system disruption is a rare event in these
clusters.

\end{abstract}


\section{Introduction} 

Current data indicates that a significant fraction of the stellar
population is born in groups and clusters embedded within the densest
regions of giant molecular clouds (GMCs).  Advances in infrared
astronomy during the past two decades have afforded astronomers with
an unprecedented view of these stellar nurseries. These clouds form
relatively rapidly ($1-10$ Myr) out of intergalactic gas and dust as a
result of the complex (and poorly understood) interplay of spiral
density waves, supernova explosions, phase transitions, and
instabilities (e.g., Elmegreen 1991).  Once formed, GMCs obtain a
highly clumpy structure, possibly due to collisions in supersonic
turbulent flows (e.g., Klessen, Heitsch \& Mac Low 2000).  This highly
nonuniform structure contains numerous cores with masses ranging from
a few to a few thousand solar masses.  These dense cores (which have
been mapped in NH$_3$ -- see the compilation of Jijina et al. 1999)
are the sites of star formation. Specifically, fragmentation within
the more massive cores ($M > 50 M_\odot$), possibly resulting from
Jeans instability, decoupling of fluid and MHD waves (Myers 1998),
and/or from the decay of turbulence (Klessen \& Burkett 2000, 2001),
form gravitationally unstable substructures whose subsequent collapse
leads to the formation of protostars (e.g., Shu 1977; Fatuzzo et al.
2004).  At the end of this complex process, young embedded 
groups/clusters appear to be basic units of star formation, accounting
for a significant fraction (perhaps as high as 90\%) of the stars that
populate our galactic disk.  The evolution of these young clusters and
their resulting effects on stellar and planetary formation represents
a fundamental set of astrophysical problems.

\begin{figure}
\figurenum{1}
{\centerline{\epsscale{0.90} \plotone{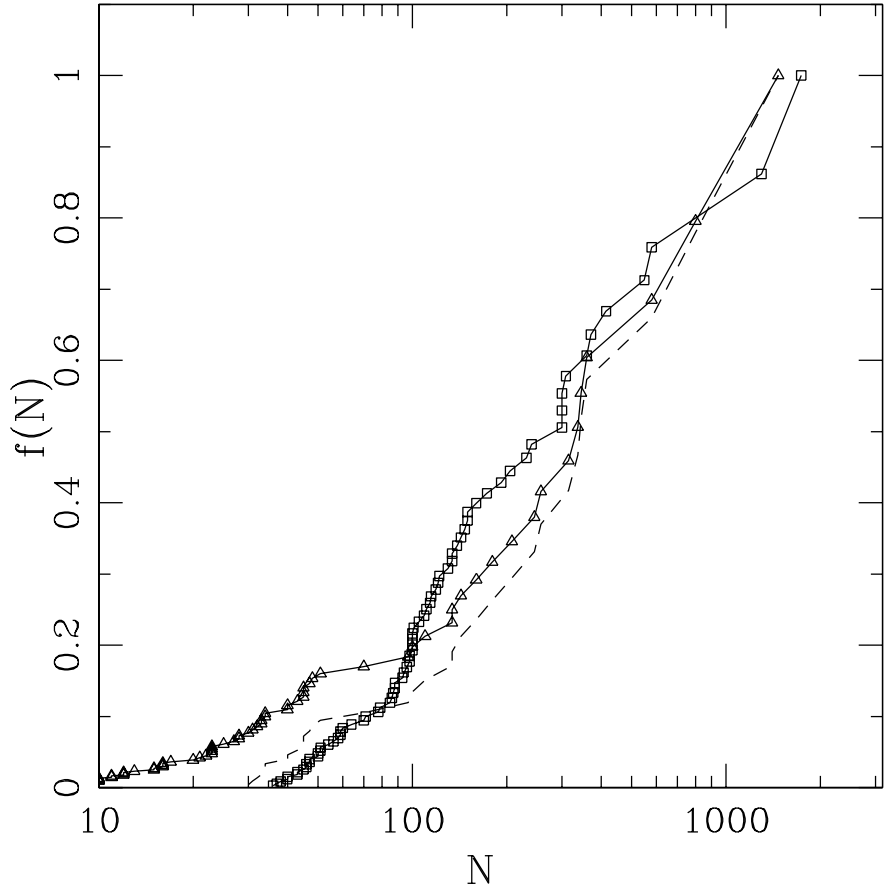} }}
\figcaption{Cumulative distribution of group/cluster sizes as a
function of system size $N$. The quantity $f(N)$ is the fraction of
the total number of stars in the sample that live in groups/clusters
of system size $N$ or smaller. The curve marked by open squares
corresponds to the 2 kpc sample, which is complete down to $N$ = 30
(Lada \& Lada 2003); the curve marked by open triangles is the 1 kpc
sample, which is complete down to $N$ = 10 (Porras et al. 2003). The 
dashed curve shows the 1 kpc sample subjected to the same selection 
criteria as the 2 kpc sample. } 
\label{fig:probn} 
\end{figure}

The ``typical'' size of star formation aggregates remains poorly
defined. Lada \& Lada (2003) and Porras et al. (2003) present catalogs
of nearby embedded clusters, the former including systems with $N \ge
30$ out to 2 kpc, and the latter including systems with $N \ge 10$ out
to 1 kpc. The cumulative distributions for the number of stars born in
units of size $N$, as a function of $N$, are presented in Figure
\ref{fig:probn} for both catalogs. The open squares represent the 2
kpc sample and the open triangles represent the 1 kpc sample; the
dashed curve shows the 1 kpc sample subjected to the same criteria as
the 2 kpc sample ($N \ge 30$).  The two samples provide a consistent
estimate for the probability distribution of group/cluster sizes. One
should keep in mind that these samples are not complete. Some of the
distant groups/clusters in the sample may have larger stellar
membership (than reported) because the faint (low mass) end of the
stellar IMF is not fully observed. On the other hand, small groups
with $N \sim 30 - 100$ may well exist and not be included in the
samples at all. As a result, the true distribution of cluster sizes
$N$ could be skewed toward either higher or lower $N$ than shown in
Figure \ref{fig:probn}. For the sake of definiteness, however, in this
paper we take this sample to be representative.

Large clusters like the Trapezium in Orion (with $N > 1000$) are 
known to be disruptive to the star formation process (e.g., St\"orzer
\& Hollenbach 1999).  In contrast, small groups with $N  \le 100$ often
have relatively little impact (e.g., Adams \& Myers 2001). As shown in
Figure \ref{fig:probn}, however, the majority of stars observed in
embedded clusters are found in systems that contain between 100 and
1000 members (at least for these observational samples).  The
evolution of these intermediate-sized systems and their effects on
star and planetary formation are thus of fundamental importance.
These systems can influence star and planet formation through
dynamical interactions among kernels, competitive accretion,
scattering interactions among star-disk systems and/or or early
planetary systems, and by disruptive radiation from other stars
(especially the larger ones that live near cluster centers).


\begin{figure}
\figurenum{2}
{\centerline{\epsscale{0.90} \plotone{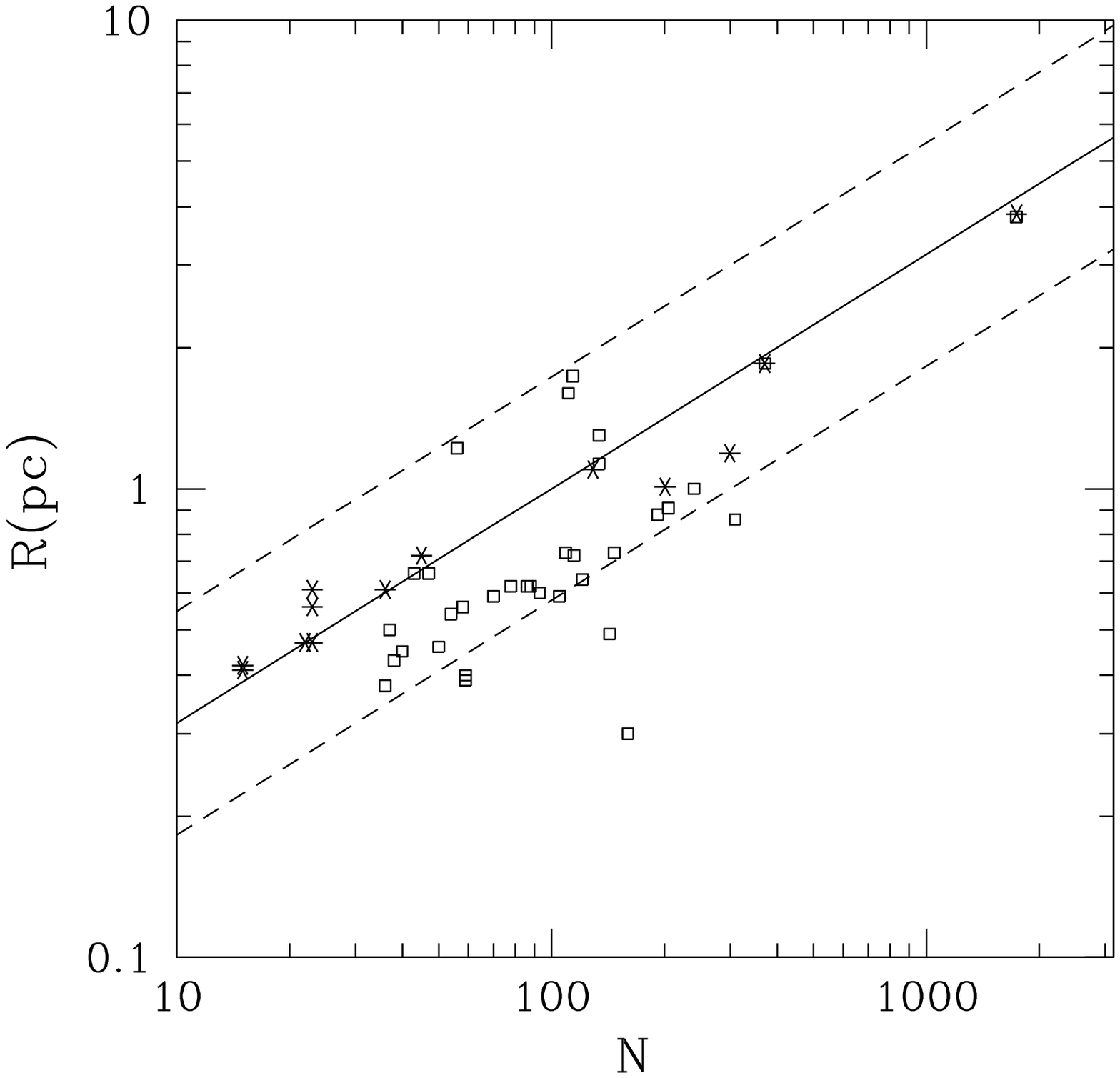} }}
\figcaption{Observed group/cluster sizes as a function of system size
$N$. The open squares represent data taken from the compilation of
Lada \& Lada (2003); the stars represent data from the work of
Carpenter (2000). The solid curve is a rough fit to the data with the
form $R(N) = R_{300} (N/300)^{1/2}$ with $R_{300}$ = $\sqrt{3}$ pc; 
the two dashed curves have the same functional dependence with the 
length scale $R_{300}$ larger or smaller by a factor of $\sqrt{3}$.  
For most of this work we use the lower curve, with $R$ = 1 pc 
$(N/300)^{1/2}$ in order to determine the greatest possible effects 
of the cluster environment. }  
\label{fig:rvsn} 
\end{figure}

This paper considers the dynamics of intermediate-sized stellar
systems with $N = 100 - 1000$.  In the two data sets described above
(Lada \& Lada 2003; Porras et al. 2003), the fraction of stars that
are found in systems with $N < 100$ is 19\% and 20\%, respectively,
whereas the fraction of stars found in systems with $N > 1000$ is 24\%
and 20\%. The majority of stars (about 60\%) are found in systems
within our range of study.  A large body of previous work on the
dynamical evolution of $N$-body systems exists. The evolution of
stellar clusters has been investigated for both small $N \le 100$
(Lada, Margulis \& Deardorn 1984) and large $N > 10,000$ (e.g.,
Portegies Zwart et al. 1998; Boily \& Kroupa 2003b).  The dynamical
effects of binaries has also been explored both in the context of
globular clusters (Hut \& Bahcall 1983) and young clusters (Kroupa,
Petr, \& McCaughrean 1999; Kroupa \& Bouvier 2003). Some work on
intermediate sized systems has been performed (see Kroupa 1995b and
references therein). On a smaller system scale, planetary disruption
has been explored by numerous authors (e.g. de la Fuente Marcos \& de
la Fuente Marcos 1997, 1999; Adams \& Laughlin 2001; Smith \& Bonnell
2001; Hurley \& Shara 2002; David et al. 2003; Fregeau, Chatterjee, \&
Rasio 2005), and the few-body problem has been investigated by Sterzik
\& Durisen (1998).  Although these works have greatly advanced our
understanding of the dynamics of many-body systems, a great deal of
work remains to be done. This paper concentrates on the range of
parameter space subtended by most young stellar clusters, $N$ in the
range $100 \le N \le 1000$, and seeks to determine the effect of the
cluster environment on forming stars and planetary systems.

In addition to its focus on intermediate sized clusters, this work
differs from previous studies in the starting conditions.  Most
previous $N$-body simulations of stellar groups have invoked virial
arguments to set the initial velocities of the system members. A
distinguishing aspect of this study is the adoption of subvirial
starting conditions. This initial condition is motivated by clump
dispersion measurements obtained from recent observations of four
systems in which the stars are (apparently) born with speeds
substantially lower than virial (assuming that observed clumps are
progenitors of individual protostars or stars). Specifically, in the
NGC 1333 cluster, the observed clump-to-clump RMS velocity is only
$\sim 0.45$ km/s, significantly less than that expected if the clumps
were in virialized orbits ($v\sim$ 1 km/s); furthermore, the
clump-to-clump RMS velocity is much lower for subgroups within the
larger complex (Walsh et al. 2004, Walsh, Myers, \& Burton
2004). Similarly, the velocity dispersion for 45 clumps
(condensations) in $\rho$ Oph was estimated to be $\sim 0.64$ km/s,
with similar results obtained for 25 clumps in the NGC 2068
protocluster (see Andr\'e 2002 and references therein). As another
example, the clump to clump velocities in the NGC 2264 region are
estimated to be about 3 times smaller than that expected in virial
equilibrium (Peretto, Andr{\'e}, \& Belloche 2005).

This paper undertakes a statistically comprehensive study of the
dynamical evolution of young stellar clusters with populations in the
range $100 \le N \le 1000$ and uses $N$-body simulations to follow
these systems from their nascent, embedded stages out to ages of 10
Myr. One goal of this study is to explore how early evolution depends
upon the number $N$ of system members. The systems begin with a
gaseous component which is subsequently removed (e.g., at time 5 Myr).
Multiple realizations of equivalent initial conditions are performed
in order to build up robust distributions of the output measures. We
find that 100 realizations (simulations) of each set of initial
conditions are required to provide good statistics for the output
measures. These measures include closest approaches of cluster
members, their radial locations and mass profiles (which largely
determine the radiation exposure), as well as the time evolution of
the bound cluster fraction, the virial ratio, the velocity isotropy
parameter, and the half-mass radius. Because of the large number of
simulations required for each set of initial conditions, we limit this
preliminary study to six cluster types: $N$ = 100, 300, and 1000, with
both ``cold'' (subvirial) and virial initial conditions.

The output measures are used to determine the impact of the cluster
environment on star and planet formation.  Toward that end, we
determine the distribution of FUV luminosities for groups and clusters
as a function of system size $N$. This ultraviolet radiation acts to
destroy circumstellar disks and to inhibit planet formation. This work
provides a measure of its efficacy as a function of group/cluster size
$N$ (\S 3). We also calculate the cross sections for newborn planetary
systems to be disrupted by passing stars (binaries). These cross
sections (\S 4) are used in conjunction with the distributions of
closest approaches from the $N$-body simulations to provide a measure
of solar system disruption as a function of system size $N$.  Armed
with a robust statistical description of the evolution of young
clusters, we undertake a detailed analysis of the particular system
NGC 1333 (\S 5).  Recent observations of this young cluster (Walsh et
al. 2005) provide position and velocity information on the $N$ = 93
N$_2$H$^+$ clumps found within the system.  Since the observations
specify only three of the six components of phase space, we must
reconstruct the cluster conditions through multiple realizations,
thereby producing an ensemble of calculations that can then be
compared with the results of our theoretical study. Our results and
conclusions are summarized in \S 6.

\section{Numerical Simulations of Young Embedded Clusters} 

For the first part of this study we perform a suite of $N$-body
simulations for intermediate-sized clusters as they evolve from their
embedded stage out to ages of $\sim 10$ Myr. Cluster evolution depends
on the cluster size $N$, the initial stellar profile, the initial gas
profile, the star formation history, the stellar IMF, and the gas
disruption history.  Given the large number of parameters needed to
adequately describe young clusters (see also below), this initial
study does not consider every combination of parameters that these
systems could attain.  Instead, we identify a baseline set of
parameters that represent a ``typical'' cluster and perform many
realizations of this benchmark model. We find that for every set of
cluster parameters, one must perform many realizations of the initial
conditions in order to fully sample the output measures.  This study
explores the variation of the cluster size $N$ and the effects of
subvirial versus virial starting conditions. A forthcoming follow-up
study will consider a much wider exploration of parameter space. For
each set of input parameters, we perform 100 equivalent realizations
in order to build up a statistical representation of the output
measures. The input parameters and output measures are described
below.

The $N$-body integrations are performed using NBODY2 (Aarseth 1999,
2001).  This version of the integration package is relatively fast and
allows for many realizations of each set of initial conditions to be
run, as required to obtain good statistics. In this initial study,
however, we do not include the binarity of the stellar systems. In
sufficiently dense and long-lived clusters, binaries can absorb and
store enough energy to affect the evolution of the cluster system.
This paper focuses on the dynamics of systems with $N$ = 100 -- 1000,
where we expect interactions to be sufficiently rare and sufficiently
distant that binarity has only a small effect on overall energy budget
of the cluster (see also Kroupa 1995, Kroupa et al. 2003).  This
approximation is checked for consistency in two ways.  First, we
perform a test simulation including binaries (using NBODY6; Aarseth
1999) and find that the results are the same. As a second check, we
use the distributions of closest approaches found from our ensemble of
simulations and find that binary interactions are not energetically
important in these systems (see below).

\subsection{Parameter Space}
 
{\it Cluster membership $N$.} Figure \ref{fig:probn} indicates that
most stars form in clusters with stellar membership $N$ in the range
$100 \le N \le 1000$, with roughly half of stars belonging to clusters
with size $N < 300$ (and half with $N>300$). We thus consider the
value $N = 300$ as the center of our parameter space, and explore the
evolution of clusters with $N$ = 100, 300, and 1000.

{\it Initial cluster radius $\rstar$.} Young clusters are found to
have radii $\rstar$ within the range 0.1 -- 2 pc. An observationally
determined relation between $\rstar$ and $N$ is shown in Figure
\ref{fig:rvsn}, where open squares represent data taken from the
compilation of Lada \& Lada (2003) and stars represent data from
Carpenter (2000).  A correlation between $\rstar$ and $N$ is clearly
evident, although significant scatter exists. The data can be fit by
the relation of the form
\be
\rstar(N) = R_{300} \sqrt{(N/300)} \, , 
\label{eq:rstar} 
\ee
where $R_{300} \approx 1 - 2$ pc.  This relation corresponds to a
nearly constant surface density of stars $N/R^2 \approx$ {\sl
constant}.  The solid curve shown in Figure \ref{fig:rvsn} uses
$R_{300}$ = 1.7 pc; the dashed curves have the same functional
dependence but are scaled (up or down) by a factor of $\sqrt{3}$, and
quantify the spread in this correlation.  For this study we adopt this
functional dependence to specify the initial radius of the stellar
component and use $R_{300} = 1.0$ pc. This value is near the lower end
of the observed range and thus maximizes the density, which in turn
leads to dynamical interactions near the upper end of the range
expected in these cluster systems.

{\it Initial stellar profile.} Embedded clusters display structure
that can be characterized as centrally condensed or hierarchical (Lada
\& Lada 2003).  In a complete treatment, one should explore both
spherical and nonspherical stellar distributions.  In this initial
study, however, we focus on the spherical case, where stars are
randomly placed within a sphere of radius $\rstar$. For the sake of
definiteness, the initial density of stars is taken to have the form
$\rho_\ast \sim r^{-1}$ so that the initial stellar mass component is
distributed according to $M_\ast(r) \sim r^{2}$ (out to the boundary at
$\rstar$). This form is consistent with the expected density profiles
for gas in cluster forming cores (see below). 

Although there is evidence for a nearly universal initial mass
function (IMF) for stars in young clusters, it remains unclear how
stellar mass correlates with the initial position within a cluster.
Massive stars are preferentially found near the centers of open
clusters (e.g., Elmegreen et al. 2000), but the same trend need not be
universally true for embedded clusters. Some clusters show evidence
for mass segregation (Testi et al. 1998; Hillenbrand \& Hartmann 1998;
Jiang et al. 2002) and theoretical considerations suggest that mass
segregation has a primordial origin in some systems (Bonnell \& Davies
1998; Hillenbrand \& Hartmann 1998; see also Carpenter et al. 1997).
However, the relative importance of dynamical versus primordial mass
segregation in clusters with $100 < N < 1000$ remains uncertain.  Given
the evidence for some primordial mass segregation, we adopt a simple
algorithm consistent with observed groups: For a given system, we
sample the stellar masses from a standard IMF, and then relocate the
most massive member to the cluster center. The remaining stars are
then placed randomly so that the initial stellar component has density
$\rho_\ast \sim r^{-1}$ within the radial range $0 \le r \le \rstar$. 
This approach thus provides a minimal treatment of primordial
mass segregation. A more detailed treatment should be considered in
follow-up studies. The issue of mass segregation is important because
massive stars can produce powerful winds, outflows, and radiation
fields that, if centralized, can more readily disrupt the gaseous
component of a cluster (as well as planet forming disks around other
stars).

{\it Initial speeds.}  As discussed above, stars often appear to be
born in young embedded clusters with initial speeds substantially less
than the virial values (Andr{\'e} 2002, Walsh et al. 2004, Peretto et
al. 2005). To set the initial stellar velocities, we sample from a
distribution that is characterized by a given expectation value for
the virial ratio $Q \equiv |K/W|$ (Aarseth 2003), i.e., the ratio of
kinetic to potential energy, where $Q$ = 0.5 for virialized
systems. One goal of this study is to explore the effects of subvirial
starting conditions. For the sake of definiteness, we adopt a baseline
value of $Q = 0.04$ for our subvirial simulations (i.e., starting
speeds about 30\% of the value needed for virial equilibrium). For
comparison, we also study the virilized initial condition $Q$ = 0.5
for (otherwise) the same starting conditions. We refer to the
subvirial starting states as ``cold'' and the virial initial states as
``virial''.

{\it Spread in star formation times.}  A system of stars evolving from
such an initially ``cold'' state would collapse into a dense core
within a crossing time if all of the stars formed (and hence began
falling toward the center) at exactly the same time. The resulting
traffic jam at the cluster center would be unphysical, however,
because the stars must have a spread in formation time.  In this
study, we assume that forming stars are tied to their kernels (the
collapsing pockets of gas), which are moving subsonically, until the
collapse phase of an individual star formation event is completed.
After their collapse phase, newly formed stars are free to fall
through the gravitational potential of the group/cluster system. Here
we assume that the star formation epoch lasts for a given span of time
$\Delta t$ = 1 Myr, which is comparable to the crossing time.  For
comparison, the expected collapse time for an individual protostar is
much smaller, only about 0.1 Myr (see Shu 1977; Adams \& Fatuzzo 1996;
Myers \& Fuller 1993). 

{\it Initial gas potential.} Observations of young embedded clusters
indicate that the gas density profiles may have (roughly) the form
$\rho \sim r^{-1}$ (Larson 1985, Myers \& Fuller 1993, Jijina et
al. 1999; see also the discussion of McKee \& Tan 2003) on the radial
scale of the cluster ($\sim 1$ pc).  For these simulations we need to
include the graviational potential of the gaseous component and
eventually let it disappear with time. In order to smoothly extend the
initial gas potential out to large radii, we adopt a Hernquist profile
so that the initial gas distribution is characterized by the
potential, density, and mass profiles of the forms
\be 
\Psi = {2 \pi G \rho_0 r_s^2 \over 1 + \xi } \, , \qquad 
\rho = {\rho_0 \over \xi (1 + \xi)^3} \, , \qquad {\rm and} 
\qquad M = { M_\infty \, \xi^2 \over (1+\xi)^2} \, ,
\label{eq:hqprofile} 
\ee 
where $\xi \equiv r/r_s$ and $r_s$ is a scale length (Hernquist 1990).
Notice that $M_\infty$ = $2 \pi r_s^3 \rho_0$.  In practice we
identify the scale $r_s$ with the cluster size (Fig. \ref{fig:rvsn}), 
so that $r_s = \rstar$. The density profile within the cluster itself
thus has the form $\rho \sim r^{-1}$; the steeper density dependence
$\rho \sim r^{-4}$ occurs only at large radii (effectively outside the
cluster) and allows the potential to smoothly join onto a force-free
background.  The mass enclosed within $\xi = 1$, denoted here as
$M_1$, is the effective gas mass within the cluster region itself
(notice that the density and mass profiles extend out to spatial
infinity and that the asymptotic mass $M_\infty = 4 M_1$). The star
formation efficiency (SFE) within the cluster is thus given by SFE =
$M_\ast / (M_\ast + M_1)$. Although observational determinations of
SFE are subject to both uncertianties and system-to-system variations,
typical values for a sample of nearby embedded clusters lie in the
range SFE = 0.1 -- 0.3 (Lada \& Lada 2003).  This study adopts a
baseline value $M_1 = 2 M_\ast$ (so that SFE = 0.33). Thus, the 
mass that will end up in stars over the time interval $\Delta t$ = 1
Myr is pre-determined. Over the time $\Delta t$, the stellar masses
become dynamically active and begin to fall through the potential
(thus, the total mass of the cluster is kept constant over the time 
$\Delta t$ when stars are being formed). 

{\it Gas removal history.} Stellar aggregates are initially deeply
embedded in dense gas, but they quickly disrupt the gaseous component
through the action of stellar winds and outflows, radiative processes,
and supernovae (e.g., Whitworth 1979; Matzner \& McKee 2000; Gutermuth
et al. 2004). Although the details of the gas removal processes are
not fully understood, observations indicate that clusters older than
about 5 Myr are rarely associated with molecular gas, so that gas
removal must occur in these systems on a comparable timescale (Lada \&
Lada 2003).  The fraction of stars that remain gravitationally bound
after gas removal has been explored both analytically (e.g., Adams
2000; Boily \& Kroupa 2003a) and numerically (e.g., Lada, Margulis \&
Dearborn 1984; Geyer \& Burkert 2001; Boily \& Kroupa 2003b).  Gas
affects the dynamical evolution through its contribution to the
gravitational potential. As gas leaves the system, the gravitational
well grows less deep and the stellar system adjusts its structure.
Stars filling the high velocity part of the distribution will thus
leave the system, but a fraction of stars can remain bound after the
gas has been removed.  The value of this fraction depends on the star
formation efficiency, the geometries of the gaseous and stellar
components, the gas dispersal history, and the stellar distribution
function. This paper uses a simple model for gas removal: The gas is
removed instantaneously at a given time $t$ = 5 Myr (e.g., Leisawitz,
Bash, \& Thaddeus 1989) after the star formation process begins
(recall that stars are randomly introduced over a time interval
$\Delta t$, the beginning of which defines the time $t$ = 0). This
choice of parameters allows the gas to remain in the system as long as
possible (according to the currently available observations -- see
Lada \& Lada 2003). These simulations thus represent an upper limit on
the level of interactions expected in astronomical clusters. Notice
also that the gas potential is considered fixed while gas remains
within the cluster.  For clusters with cold starting conditions, the
stars fall toward the cluster center and the gas could become more
concentrated as well. This effect is small in the present case because
gas dominates the potential, but could be considered in further work.

{\it Binary test.} In order to test the validity of our approximation
of ignoring binarity, we performed a test simulation using both NBODY6
(which includes binaries -- Aarseth 1999) and NBODY2 (where the masses
of the two binary companions are combined to make a single star). The
comparison runs are made for a cluster with $N$ = 300 and radius $R$ =
1 pc, which defines the center of our parameter space (see above). We
also use a cold start, an initial $Q$ = 0.04, because the cold runs
should have more interactions and hence be more affected by binaries.
In the test runs, gas is included as a Plummer sphere (with scale
radius $R_s$ = 1 pc) since the original $N$-body codes are written
with the Plummer potential. The gas mass is equal to the total stellar
mass.  Over a time scale of 10 Myr, we find that the evolution of the
fraction $f_b$ of bound stars, the virial parameter $Q$, and the
half-mass radius $R_{1/2}$ are virtually identical for the two cases.

\subsection{Output Measures}

One goal of this work is to provide a statistical description of the
systems under study. Two systems with identical sets of cluster
parameters ($N$, $\rstar$, \dots) will have stars located at different
starting locations and can evolve in different ways (for example, the
history of close encounters will change).  To provide a more complete
description of the evolution of young clusters, we perform an ensemble
of ``effectively equivalent'' simulations through multiple
realizations of the system, i.e., we use the same set of cluster
parameters but different choices for the random variables.  In this
manner, we can build up full distributions for the output measures of
the systems (see also Goodman Heggie, \& Hut 1993; Gierrsz \& Heggie
1994; Baumgardt, Hut, \& Heggie 2002; and references therein). 

\subsubsection{Time evolution} 

To characterize the time evolution, a variety of output measures are
computed for each simulation, including the cluster's bound fraction,
virial ratio, half-mass radius, and velocity isotropy parameter.
These measures are calculated every 0.25 Myr throughout each 10 Myr
simulation.  For each system studied, the output measures of all the
realizations (effectively equivalent simulations) are combined and
averaged.  We can then investigate the temporal evolution of each
measure as well as use the measures to compare the different systems
studied.

\begin{figure}
\figurenum{3} 
\centerline{\epsscale{0.80} \plotone{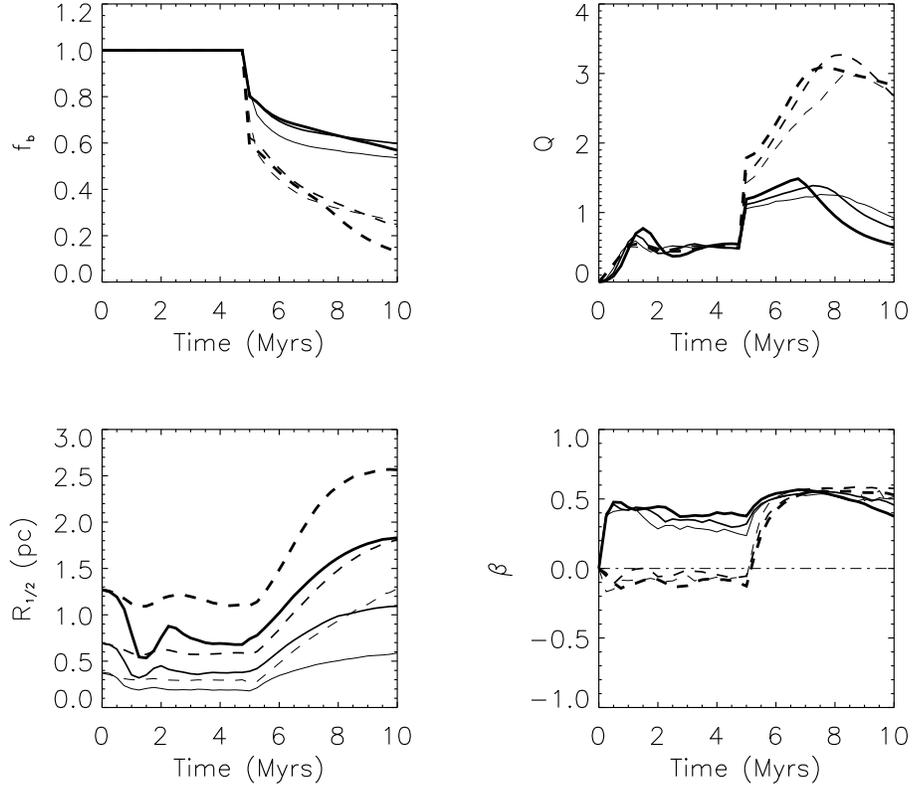} }
\vskip -2.2truein
\figcaption{Time evolution for output measures for cluster simulations 
with $N$ = 100, 300, and 1000.  In each panel, the solid curves show
the time evolution functions for ``cold'' initial conditions and the
dashed curves show the time evolution for ``virial'' starting conditions.
The boldness of the curves denotes the cluster size, with the darkest 
curves for $N$ = 1000 and the lightest curves for $N$ = 100.
The upper left panel shows the fraction of stars that are bound as a
function of time, the upper right panel shows the time evolution of
the virial ratio $Q \equiv |K/W|$, the lower left panel shows the time
evolution of the half-mass radius, and the lower right panel shows the
velocity isotropy parameter $\beta$. In all of these simulations,
the gas is removed at time $t$ = 5 Myr, which leads to structure in
all of the time evolution functions as shown here. }
\label{fig:output} 
\end{figure}

One important quantity is the fraction $f_b$ of stars that remain
gravitationally bound as a function of time. For example, we would
like to know how $f_b (t)$ depends upon the cluster size $N$ and the
starting conditions (virial versus cold).  The bound fraction $f_b$ of
the cluster is defined by $f_b \equiv N_{bound}/N$, where $N$ is the
initial number of stars in the cluster and $N_{bound}$ is the number
of stars which have negative total energy at a given time.  The bound
fraction functions $f_b(t)$ are shown in Figure \ref{fig:output} for
the six types of clusters considered here. Gas is removed at time $t$
= 5 Myr, so the fraction of bound stars decreases after that
epoch. Figure \ref{fig:output} shows that the cold clusters retain
more of their stars for longer times.

In addition to $f_b$, we track the evolution of three other cluster
diagnostics.  The half-mass radius $R_{1/2}$ is defined to be the
radius that encloses half of the stellar mass that is still
gravitationally bound to the system. Over the long term, the half-mass
radius $R_{1/2}$ is an increasing function of time, although it can
decrease during the initial evolution of cold clusters.  Within
groups/clusters, young stars are often born with speeds much smaller
than that required for virial equilibrium, but attain larger (virial)
speeds as the system evolves. We can monitor this approach to
equilibrium by tracking the evolution of the virial ratio $Q$ (the
ratio of kinetic to potential energy for the stellar population).  We
also track the evolution of the isotropy parameter $\beta \equiv 1 -
v_\theta^2 / v_r^2$, where $v_\theta$ and $v_r$ are the (averaged) 
$\hat \theta$ and $\hat r$ components of the velocity.  The isotropy
parameter provides a measure of the degree to which the cluster
members have radial orbits.  An isotropy parameter of $\beta = 0$
corresponds to an isotropic velocity distribution, whereas $\beta = 1$
corresponds to a cluster where the members are moving primarily in the
radial direction.


The time evolution of the aformentioned cluster diagnostics are shown
in Figure \ref{fig:output} for cluster sizes $N$ = 100, 300, 1000, and
for both ``cold'' and ``virial'' starting conditions. As shown, the
bound fraction is a slowly decreasing function of time, with a
substantial jump at $t$ = 5 Myr when the gas is removed. The half mass
radius $R_{1/2}$ remains nearly constant until gas removal at $t$ = 5
Myr, when it becomes an increasing function of time.  The isotropy
parameter $\beta$ is substantially radial ($0 < \beta < 1$) over the
entire evolution time for clusters with cold starting conditions, but
shows a slight downward tendency at late times, indicating some
evolution towards isotropy. For virial clusters, the parameter $\beta$
is close to zero (isotropic) for the first 5 Myr of evolution, but
develops a definite radial characteristic ($\beta \sim 0.5$) for the
second half of the time interval after the gas is removed.  These same
general trends are evident in the ensemble of results summarized by
Table 1 below. For each cluster size ($N$ = 100, 300, 1000) and each
starting condition (``virial'' or ``cold'') we have found the average
values of the fraction $f_b$ of stars that remain bound after 10
Myr. Similarly, we have found averages of the viral parameter $Q$, the
half-mass radius $R_{1/2}$, and the isotropy parameter $\beta$ for the
first 5 Myr (while the clusters retain gas) and the second 5 Myr of
evolution (when the clusters are gas-free). The final line of the
table gives the output parameters for our simulations of NGC 1333 (see
\S 5).

\bigskip 
\centerline{\bf Table 1: Cluster Evolution Parameters}  

\begin{center}
\begin{tabular}{lccccccc}
\hline 
\hline 
  & $f_b$ & $Q$ & $Q$ & $R_{1/2}$ (pc) & $R_{1/2}$ (pc) & $\beta$  & $\beta$  \\
Cluster Type & 10 Myr & 0-5 Myr & 5-10 Myr & 0-5 Myr & 5-10 Myr & 0-5 Myr & 5-10 Myr \\ 
\hline
100 Cold   &  0.536 & 0.489  & 1.15   & 0.211  & 0.457 &  0.320   & 0.502 \\
100 Virial    &  0.265 & 0.511  & 2.48   & 0.301  & 0.832 & --0.0849 & 0.515 \\
300 Cold   &  0.598 & 0.491  & 1.15   & 0.413  & 0.861 &  0.368   & 0.511 \\
300 Virial    &  0.239 & 0.517  & 2.72   & 0.596  & 1.31  & --0.0404 & 0.513 \\
1000 Cold  &  0.569 & 0.497  & 1.04   & 0.780  & 1.44  &  0.410   & 0.500 \\
1000 Virial   &  0.130 & 0.527  & 2.74   & 1.15   & 2.09  & --0.0993 & 0.485 \\
\hline
NGC 1333   &  0.689 & 0.525  & 0.690  & 0.117  & 0.238  & 0.230   & 0.339 \\ 
\hline 
\hline 
\end{tabular}
\end{center}

\subsubsection{Radial distributions} 

\begin{figure}
\figurenum{4} 
{\centerline{\epsscale{0.80} \plotone{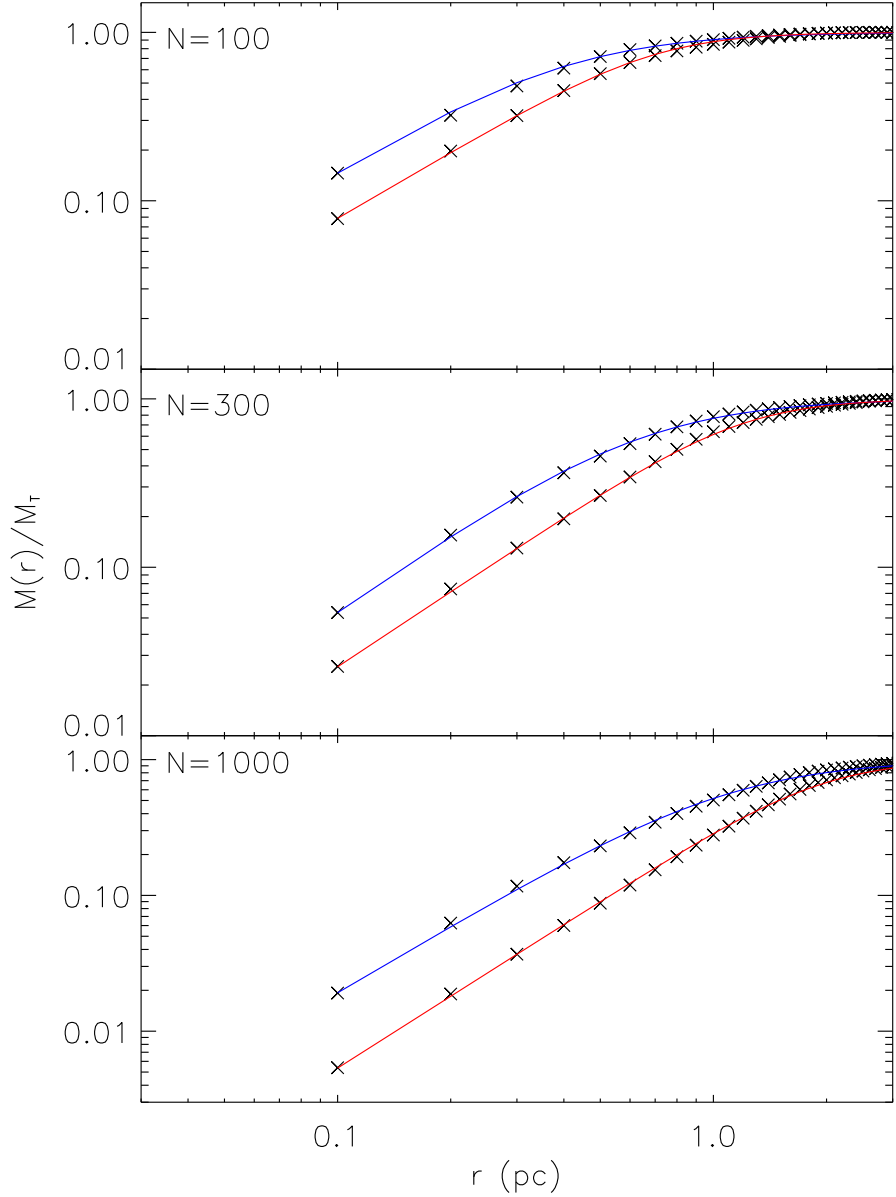}  }} 
\figcaption{Time-averaged mass profiles $M(r)/M_{T\ast}$ for the six classes
of starting conditions. The averages are taken over the first 10 Myr
and include only the stars that remain bound to the cluster (at each
time).  The top panel shows the stellar mass distribution $M(r)$ as a
function of radius $r$ for clusters with $N$ = 100 and both ``virial''
(lower curve) and ``cold'' (upper curve) starting conditions.  Each
mass profile is compiled from the results of 100 simulations with 
different realizations of the same starting conditions. Similarly, the
middle panel shows the mass distributions for clusters with $N$ = 300
and the bottom panel shows the distributions for $N$ = 1000. }
\label{fig:radial} 
\end{figure}

As a group/cluster system evolves, interactions between members result
in a distribution of stellar positions and velocities.  As the gas is
removed from the system, high-velocity stars are more likely to become
gravitationally unbound and leave the system, whereas low velocity
stars tend to condense into a central bound core. Complicating this
process, dynamical mass segregation also takes place, albeit on
somewhat longer time scales.  As one way to characterize the evolution
of these sytems, we produce mass profiles $M(r)$ averaged over the 10
Myr time interval of interest.  Specifically, the radial position of
every star is recorded at intervals of 0.25 Myr throughout each
simulation. The resulting data set is used to create a mass profile
$M(r)/M_{T\ast}$ at each time, where $M_{T\ast}$ is the total mass in
stars that remain bound. The profiles are then averaged over all time
steps and averaged over the 100 equivalent realizations of the system
to produce the radial mass profile associated with each type of
group/cluster. The integrated mass distribution $M(r)$ can be fit with
a simple function of the form
\be
{M (\xi) \over M_{T\ast}} = \left( {\xi^a \over 1 + \xi^a} 
\right)^p \, , 
\label{eq:mfit} 
\ee
where $\xi$ = $r/r_0$, and where the scale length $r_0$ and the index
$p$ are free parameters that are fit to the output of the simulations.
The index $a$ can also be varied: We find that the cold clusters can
be fit with $a$ = 2, whereas the virial clusters require $a$ = 3. The
best fit parameters for the various simulations are given in Table 2
below. The table also shows the fitting parameters for the mass
profiles averaged over the first 5 Myr (before gas removal) and over
the second 5 Myr (after gas has left the system).  The final line of
the table gives the fitting parameters for the simulations of NGC 1333
(see \S 5).  Figure \ref{fig:radial} shows the radial mass profiles
from both the simulations and the fitting functions. The simulation
profiles are time averaged over the first 10 Myr of evolution (and
over 100 realizations of each starting condition).

\newpage
\centerline{\bf Table 2: Output Parameters for Mass Distributions} 

\begin{center}
\begin{tabular}{l|c|c|c|c} 
\hline
\hline 
& (0 -- 10 Myr) & (0 -- 5 Myr) & (5 -- 10 Myr) & \\ 
Cluster Type & $p$ \qquad $r_0$ (pc) & $p$ \qquad $r_0$ (pc) & 
$p$ \qquad $r_0$ (pc) & $a$ \\
\hline
100 Cold   & 0.686 \qquad 0.394  & 0.680 \qquad 0.264 & 1.02  \qquad 0.453  & 2   \\
100 Virial    & 0.436 \qquad 0.698  & 0.406 \qquad 0.486 & 0.776 \qquad 0.793  & 3   \\
300 Cold   & 0.785 \qquad 0.635  & 0.747 \qquad 0.484 & 1.01  \qquad 0.781  & 2   \\
300 Virial    & 0.493 \qquad 1.19   & 0.489 \qquad 0.846 & 0.605 \qquad 1.56   & 3   \\
1000 Cold  & 0.820 \qquad 1.11   & 0.769 \qquad 0.899 & 0.970 \qquad 1.34   & 2   \\
1000 Virial   & 0.586 \qquad 1.96   & 0.590 \qquad 1.53  & 0.689 \qquad 2.45   & 3   \\
\hline
NGC 1333   & 0.552 \qquad 0.300  & 0.436 \qquad 0.241 & 0.924 \qquad 0.299  & 2   \\
\hline 
\hline 
\end{tabular}
\end{center}


We can also find profiles $N(r)$ for the number of stars enclosed
within the radius $r$. These profiles are essentially the probability
distributions for the radial positions of the stellar members. These
profiles $N(\xi)/N_T$ can be fit with the same form as equation
(\ref{eq:mfit}). Although not shown here, the fitting parameters are
nearly the same as those of the mass profiles and are used (in \S 3)
when we need to calculate the probability of finding a star at radius
$r$.

One goal of this study is to characterize this class of groups and
clusters. Toward this end, recent observational studies have
determined the central densities for clusters (e.g., Gutermuth et
al. 2005). However, the mass profiles found here imply that the
central denssity of these clusters suffers from an ambiguity: If a
mass profile has the form given by equation (\ref{eq:mfit}), the
density profile takes the form $\rho \propto a p \xi^{ap-3} (1 +
\xi^a)^{-(p+1)}$, which diverges in the limit $\xi \to 0$. As a
result, the central density for these mass profiles -- and this class
of systems -- is not well-defined. In contrast, the total depth of the
gravitational potential is well-defined and can be written in the form
\be 
\Psi_\ast = {G M_{T\ast} \over r_0} \psi_0 \qquad {\rm where} \qquad 
\psi_0 \equiv \int_0^\infty \Bigl( {1 \over 1 + u^a} \Bigr)^p \, , 
\ee 
where we assume that the mass profile has the form of equation
(\ref{eq:mfit}), which defines the indices $a$ and $p$, as well as 
the scale length $r_0$. The total mass of stars in the cluster is
$M_{T\ast}$, so that $\Psi_\ast$ represents the total depth of the
stellar contribution to the potential. For embedded clusters, the gas
contribution to the potential (eq. [\ref{eq:hqprofile}]) should be
added to obtain the total potential. Notice that in the limit 
$ap \to 1$, the integral in the definition of $\psi_0$ diverges and
the central potential is no longer defined. However, all of the
clusters considered here display well-defined central potentials.  For
the indices listed in Table 2, the value of the dimensionless
potential $\psi_0$ lies in the range $\psi_0$ = 1.5 -- 5.4. The
resulting stellar potential can be written in terms of a velocity
scale $\sqrt{\Psi_\ast}$, which falls in the range $\sqrt{\Psi_\ast}$
= 0.64 -- 2.4 km/s for the clusters considered here.


\subsubsection{Distribution of closest approaches} 

The cluster environment facilitates close stellar encounters which can
disrupt solar systems.  Within the ensemble of $N$-body simulations
described above, we can find the distributions of close encounters.
These distributions, in conjunction with the cross-sections for
disruptions of planetary systems (see \S 4; Adams \& Laughlin 2001),
binary-disk systems (Ostriker 1994; Heller 1993, 1995; Kobayashi \&
Ida 2001), and binary-star interactions (Heggie et al. 1996; McMillan
\& Hut 1996; Rasio et al. 1995), can then be used to estimate the
probability of interactions as a function of system size $N$ (and
other initial conditions).

Specifically, the close encounters for each star are tracked
throughout each cluster simulation; the resulting data is labeled with
both stellar mass and cluster age.  The total distribution of closest
approaches for each simulation is calculated, and these distributions
are then averaged over the 100 equivalent realizations of the system.
The result is an integrated distribution of closest approaches for
each type of cluster. The results are presented in terms of an
interaction rate, i.e., the number of close encounters with $r \leq b$
that the ``typical star'' experiences per million years (1 Myr is a
convenient unit of time and is approximately the cluster crossing
time).  This interaction rate is a function of closest approach
distance $b$ and can be fit with an expression of the form 
\be 
\Gamma = \Gamma_0 \left( {b \over 1000 {\rm AU} } 
\right)^{\gamma} \, . 
\label{eq:ratenumer} 
\ee 
The rate $\Gamma$ is thus the number of close encounters with $r \leq
b$ per star per million years.  For each type of group/cluster, the
parameters $\Gamma_0$ and $\gamma$ were varied to find the best fits
using the Levenberg-Marquardt procedure. The resulting parameter
values are given in Table 3 for the six classes of systems studied
here. The table also lists the fitting parameters for the closest
approach distributions taken over the first 5 Myr of the simulations
(when gas is still present) and the second 5 Myr time interval (after
gas removal). These results are consistent with those obtained
previously (e.g., Scally \& Clarke 2001 find similar interaction rates
for the Orion Nebula Cluster, which is somewhat larger with $N$ = 4000
stars).  Notice that the interaction rates are higher for the first 5
Myr interval than the second 5 Myr, by a factor of $\sim5$, consistent
with the spreading out of the cluster with time, especially after gas
is removed at the 5 Myr mark. Notice also that the total interaction
rate over 10 Myr is the average of the values over the two separate
time intervals. The interaction rates are higher for the clusters with
``cold'' starting conditions. In these systems, the orbits are more
radial than in the case of ``virial'' initial conditions (where the
velocity distributions are more isotropic -- see Table 1) and more
of the stars pass near the cluster center where the density is higher.

The fitting functions (given by equation [\ref{eq:ratenumer}] and
Table 3) provide a good working description of the distribution of
closest approaches for each ensemble of simulations with given
starting conditions. In order to interpret the meaning of these
results, it is useful to compare with analytic estimates (Binney \&
Tremaine 1987, hereafter BT87).  For a cluster of size $N$ and
radius $R$, the surface density of stars is roughly $N/ (\pi R^2)$.
For each crossing time $\tau_c$ of the cluster, a given star will thus
experience close encounters at the rate
\be 
\delta \Gamma_a \approx {N \over \pi R^2} 2 \pi b \delta b 
\, \tau_c^{-1} \, , 
\ee 
with impact parameter (the closest approach distance in this
approximation) between $b$ and $b + \delta b$ (BT87), where the time
unit is the crossing time. The total rate $\Gamma_a$ of close
approaches (at distance $\le b$) per crossing time is thus
approximately given by
\be
\Gamma_a \approx {N b^2 \over R^2} \tau_c^{-1} \, \approx 0.007  
\left( {b \over 1000 {\rm AU} } \right)^2 \tau_c^{-1} \, , 
\label{eq:rateanal} 
\ee 
where we have used the observed scaling of cluster radius with $N$
(Figure \ref{fig:rvsn} and equation [\ref{eq:rstar}]) to obtain the
second approximate equality. Since the cluster crossing times $\tau_c$
are of order 1 Myr, this calculation produces an interaction rate with
the same form as the fitting formula with $\Gamma_0 \approx$ 0.007 and 
$\gamma$ = 2.  As shown in Table 3, the fitting parameters for close
approaches are in rough agreement with these estimated values, at the
crudest level of comparison. The detailed forms are somewhat different
-- the numerically determined interaction rates are less steep (as a
function of $b$) and somewhat larger than the analytic estimate.

The differences between the numerical interaction rates and the
analytic estimate arise for several reasons. The crossing time is
somewhat shorter than 1 Myr, so the rate increases accordingly.  In
addition, clusters have enhanced density in their centers and support
more interactions there. Suppose that a cluster has a core, a
long-lived central region of enhanced stellar density. If the core
contains $N/10$ stars at a given time and has radius $\rstar/10$, the
effective surface density, and hence the interaction rate per crossing
time, would be 10 times higher than the estimate given above. Notice
that the local crossing time would be smaller by a factor of $\sim10$,
but that stars would (on average) spend only 10\% of their time in the
core, so these latter two effects tend to cancel.  The shallower slope
of the distribution ($\gamma < 2$) is expected for sufficiently close
encounters where the interaction energy $2Gm/b$ is comparable to the
typical velocity $V_0$ of a cluster star. In this case, the stars no
longer travel on straight-line trajectories during the encounter (as
implicitly assumed above) so that the impact parameter is larger than
the distance of closest approach (BT87).  Since $V_0 \sim 1$ km/s,
this effect comes into play when $b \sim 2 G m/V_0^2 \sim 900$ AU,
i.e., just inside the regime of interest. In the extreme limit of
small $b$, the interaction rate becomes linear $\Gamma_a \propto b
(2Gm/V_0^2)$ so that $\gamma \to 1$. Notice that the slopes found from
the numerical simulations lie in the range $1 \le \gamma \le 2$.

\newpage 
\centerline{\bf Table 3: Output Parameters for Distributions of Closest Approach} 

\begin{center}
\begin{tabular}{l|c|c|c|c} 
\hline
\hline 
& (0 -- 10 Myr) & (0 -- 5 Myr) & (5 -- 10 Myr) & \\ 
Cluster Type & $\Gamma_0$ \qquad $\gamma$ & $\Gamma_0$ \qquad $\gamma$ 
& $\Gamma_0$ \qquad $\gamma$ & $b_C$ (AU) \\ 
\hline
100 Cold   & 0.166  \qquad 1.50  & 0.266  \qquad  1.54  & 0.0672  \qquad 1.42  & 713 \\   
100 Virial    & 0.0598 \qquad 1.43  & 0.0870 \qquad  1.46  & 0.0333  \qquad 1.37 & 1430 \\
300 Cold   & 0.0957 \qquad 1.71  & 0.168  \qquad  1.73  & 0.0240  \qquad 1.59 & 1030 \\
300 Virial    & 0.0256 \qquad 1.63  & 0.0440 \qquad  1.64  & 0.00700 \qquad 1.53 & 2310 \\
1000 Cold  & 0.0724 \qquad 1.88  & 0.133  \qquad  1.89  & 0.0112  \qquad 1.83 & 1190 \\
1000 Virial   & 0.0101 \qquad 1.77  & 0.0181 \qquad  1.79  & 0.00210 \qquad 1.74 & 3650 \\
\hline 
NGC 1333   & 0.941 \qquad  1.56  & 1.39   \qquad  1.62  & 0.490   \qquad 1.42 & 238 \\  
\hline
\hline
\end{tabular}
\end{center}

\begin{figure}
\figurenum{5} 
{\centerline{\epsscale{0.80} \plotone{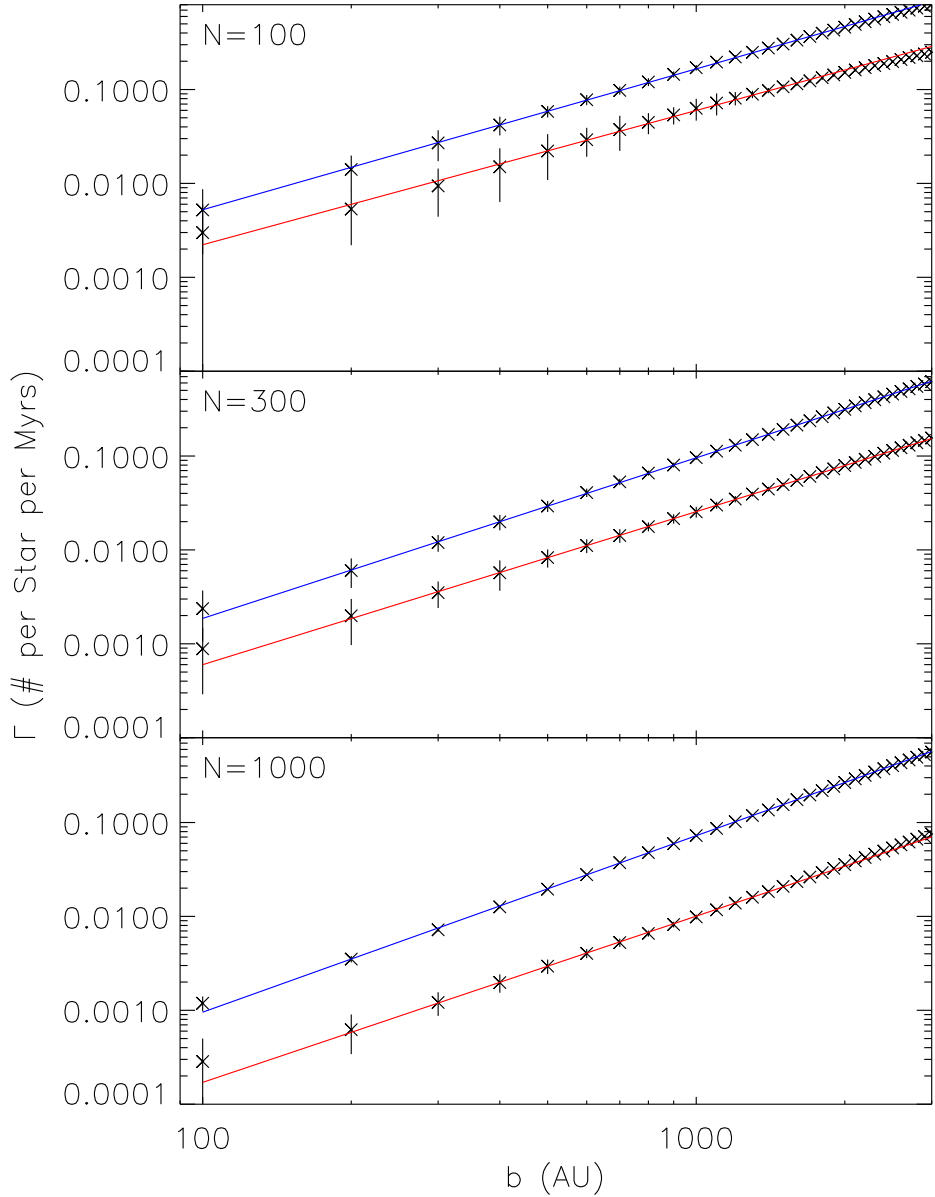}  }} 
\figcaption{Distribution of closest approaches for the six classes of 
starting conditions. The top panel shows the distributions of closest
approaches, plotted as a function of impact parameter, for clusters 
with $N$ = 100 and both ``virial'' (bottom curve) and ``cold'' (top curve) 
starting conditions.  Each distribution is compiled from the results
of 100 simulations, i.e., 100 realizations of the same starting
conditions. Similarly, the middle panel shows the distributions of
closest approaches for clusters with $N$ = 300 and the bottom panel
shows the distributions for $N$ = 1000. The error bars shown represent 
the standard deviation over the compilations.} 
\label{fig:interact}
\end{figure}

These simulations were performed using an $N$-body code that does not
consider the binarity of the interacting units (NBODY2; Aarseth 1999).
Given the distribution of closest approaches calculated here, we can
check this approximation for self-consistency. For an interaction rate
of the form of equation (\ref{eq:ratenumer}), and for a 10 Myr time
span, the ``typical'' star will experience (on average) one encounter
with the characteristic impact parameter $b_C$ given by 
\be 
b_C \equiv 1000 \, {\rm AU} \, 
\Bigl( 10 \Gamma_0 \Bigr)^{-1/\gamma} \, . 
\ee
For the six classes of clusters considered here, the characteristic
impact parameter lies in the range $b_C$ = 700 -- 4000 AU (as listed
in Table 3). For comparison, the peak of the binary period
distribution is at $P \approx 10^5$ d (Duquennoy \& Mayor 1991), which
corresponds to a separation of $\sim$ 42 AU $\ll b_C$. These results
indicate that for the majority of binary systems, the separation is
much less than the typically expected close approach $b_C$ (over the
10 Myr time span considered here).  Furthermore, the orbital energy at
the typical flyby radius of a couple thousand AU corresponds to a
velocity scale of $\sim 0.5$ km/s, i.e., a typical star is expected to
receive only a single velocity perturbation of this magnitude. 
Although this velocity kick is comparable to the mean velocity scale
of the cluster given by $v_m \sim G N \langle m \rangle / R$, most of
the interactions take place near the center of the cluster potential
where $v \sim \sqrt{\Psi_0} \sim 3.5 v_m$, so the expected velocity
perturbations are not devastating (one $\sim 15\%$ kick in velocity, 
a $\sim 2\%$ kick in energy) . These results, taken together, imply
that our approximation of ignoring binarity is justified. For
completeness we note that some binary systems can be produced via
three-body interactions during the evolution of a cluster. Care must
be taken to identify these systems once formed, and to exclude the
binary orbits from the determination of the closest approach
distribution.

\subsubsection{Comparison of virial and cold starting conditions} 

One issue of interest is the differences between the clusters with
initial conditions where members are in virial equilibrium, $Q_0 =
0.5$ (``virial'' clusters), and those where the members are started
with sub-virial velocities, $Q_0 = 0.04$ (``cold'' clusters).  The
initial conditions lead to some important differences, as illustrated
in Table 4, which lists the ratios of the parameters for the cold
simulations to those from the virial simulations.  The cold clusters
have higher bound fractions, with $50 - 60\%$ of their members
remaining bound at $t = 10$ Myr. For comparison, the virial clusters
have only $13 - 27\%$ of the initial cluster membership bound at $t =
10$ Myr. In the cold clusters, stars fall toward the center after they
form and thus spend more of their time inside the cluster gas (which
is assumed to be static, i.e., not in a state of collapse). The
removal of gas thus has less effect on the cold clusters and more
stars remain bound.  For both virial and cold clusters, the initial
velocity distribution of the cluster members is isotropic.  Since
stars in the cold clusters are started with small velocities, however,
they tend to fall toward the cluster center and rapidly develop
relatively larger radial velocities. As a result, the isotropy
parameter $\beta$ for cold clusters is larger (more radial with $\beta
\approx 0.3 - 0.4$) than that of the virial clusters, which have
$\beta \approx 0$ (and slightly negative) for the first 5 Myr of
evolution.  After the gas is removed, both the virial and cold
clusters have larger isotropy parameters, indicating increased radial
motion as the cluster expands in the absence of the gas potential.

\bigskip 
\centerline{\bf Table 4: Ratio of Parameters for Cold and Virial Initial Conditions} 

\begin{center}
\begin{tabular}{cccccc}
\hline 
\hline 
$N$  &  $f_b$ & $Q$ & $R_{1/2}$ & $R_{1/2}$ & $\Gamma_0$ \\ 
($t$ = 0) & 0--10 Myr & 5--10 Myr & 0--5 Myr  & 5--10 Myr  & 0--10 Myr \\ 
\hline
100 Stars  & 2.02 & 0.464  & 0.701  & 0.549  & 2.78 \\
300 Stars  & 2.50 & 0.423  & 0.693  & 0.657  & 3.74 \\
1000 Stars & 4.38 & 0.380  & 0.678  & 0.689  & 7.17 \\       
\hline
\hline 
\end{tabular}
\end{center}

The mass distributions (Fig. \ref{fig:radial}) and distributions of
closest approaches (Fig. \ref{fig:interact}) also depend on the starting
conditions.  The cold clusters have more centrally condensed radial
profiles throughout the 10 Myr evolution time. Before gas expulsion,
the half-mass radii for the cold clusters are about 70\% of the
half-mass radii of the virial clusters. This result can be understood if
the cold clusters act as if they have zero temperature starting
states, so they collapse to a radial scale roughly $\sqrt{2}$ times
smaller than their initial size. In comparison, virial clusters tend to
retain their starting radial size (before gas removal).  After gas
expulsion, the cold clusters continue to have half mass radii about
70\% of those of the virial clusters, although the cold clusters with $N$
= 100 are somewhat more concentrated.  The mass distributions show
that the scale radii of the cold clusters are 55--65\% of those of the
virial clusters, and the central potentials are deeper by a factor of
$1.2 - 2$.  The distributions of closest approach for clusters with
virial and cold starting conditions are similar and can be fit with a
single power-law form over the radial range of interest (100 -- 1000
AU). The power-law indices are roughly the same, but the cold clusters
show a higher interaction rate by a factor of 3 -- 7.

        
\section{Effects of Cluster Radiation on Forming Solar Systems} 

Given the characterization of cluster dynamics found in the previous
section, we now estimate the effects of ultraviolet radiation from
the background cluster on nascent solar systems. The radiation from 
the parent star can drive mass loss from its planet-forming disk and
thereby affect planet formation (Shu, Johnstone, \& Hollenbach 1993),
but this effect is often dominated by radiation from the background
star cluster (Johnstone, Hollenbach, \& Bally 1998; Adams \& Myers
2001). In this paper we focus on the effects of FUV radiation (Adams
et al. 2004), which tends to dominate the effects of EUV radiation
(Armitage 2000).  We first calculate the distributions of FUV
luminosity (\S 3.1) and then determine the extent to which the
radiation compromises planet formation (\S 3.2). Throughout this
section, we present results as a function of cluster size $N$. In this
context, we consider the cluster to have a stellar membership of $N$
primaries and we ignore binarity. Although some fraction of the
cluster members will have binary companions, the vast majority of the
companions will have low mass and will not contribute appreciably to
the FUV luminosity. Notice that in the early stages of cluster
evolution, the system will contain a substantial amout of gas and
dust, and this dust can attenuate the FUV radiation. The gas (and
dust) is removed from the cluster relatively early so that the FUV
radiation eventually has a clear line of sight to affect forming
planetary systems. Nonetheless, since we do not model the dust
attenuation, the results of this section represent an upper limit to
the effects of radiation.

\subsection{Probability Distributions for FUV Luminosity} 

In this subsection, we calculate probability distributions for the FUV
radiation emitted by stellar aggregates with varing size $N$ (the number
of cluster members).  The total FUV luminosity from a cluster or group
of stars is given by the sum
\be
L_{FUV} (N) = \sum_{j=1}^N L_{FUVj} \, , 
\label{eq:luvsum} 
\ee
where $L_{FUVj}$ is the FUV luminosity from the $jth$ member. 
In this approximation, we assume that the FUV luminosity for 
a given star is determined solely by the stellar mass. We further 
assume that the stellar mass is drawn from a known probability 
distribution, i.e., a known stellar initial mass function. 

In this problem, low mass stars have a negligible contribution to the
total UV flux. To a good approximation, we can ignore the contribution
of all stars smaller than 1 $M_\odot$.  To specify the initial mass
function, we are thus not concerned with the detailed shape at low
stellar masses; we only need to correctly account for the fraction of 
stars with $M_\ast > 1 M_\odot$ and the slope at high stellar masses. 
For the sake of definiteness, we assume that the stellar IMF has a 
power-law form for mass $M_\ast > 1 M_\odot$ with index $\Gamma$, i.e., 
\be
{dN \over dm} = A m^{-\Gamma} \, = \fone (\Gamma - 1) m^{-\Gamma} \, , 
\ee
where $m$ is the mass in units of solar masses and where the slope
$\Gamma = 2.35$ for the classic form first suggested by Salpeter
(1955); this slope remains valid today for a wide variety of regions  
(Massey 2003 and references therein).  In the second equality, we 
have normalized the distribution according to the convention 
\be
\int_1^\infty {d N \over d m} dm = \fone \, , 
\ee
where $\fone$ is the fraction of the stellar population with mass
larger than 1 $M_\odot$.  For a typical stellar mass function (such as
that advocated by Adams \& Fatuzzo 1996), the fraction $\fone \approx
0.12$. For completeness, note that in practice we cut off the IMF at
$m$ = 100 (see below); taking the integral out to $\infty$ here
results in an inaccuracy of $\sim$0.2\% (which is much smaller than
the accuracy to which we know $\fone$).

\begin{figure}
\figurenum{6}
{\centerline{\epsscale{0.90} \plotone{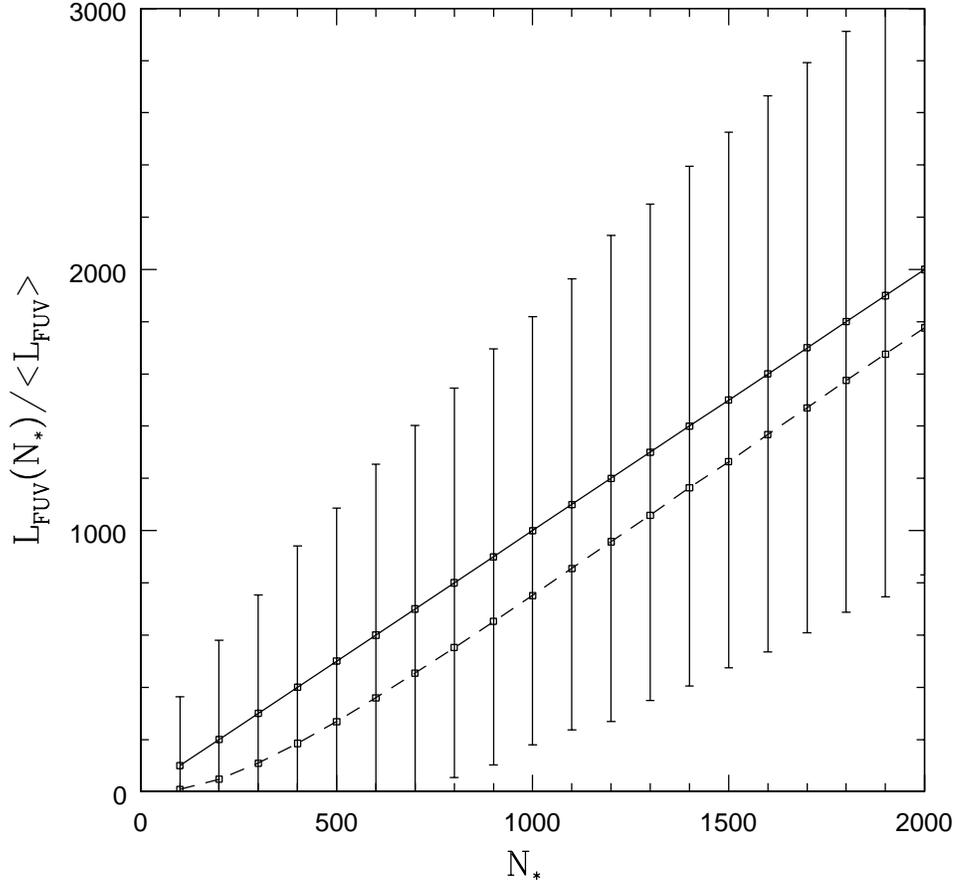}  }}
\figcaption{Mean, variance, and median of the distribution of total
FUV luminosity for clusters of size $N$ as a function of $N$. The open
square symbols connected by the solid line depict the mean FUV
luminosity averaged over many realizations of a cluster of size $N$.
The error bar symbols represent the variance about the mean. As shown
in the text, the mean $\langle L_{FUV} \rangle \propto N$ and the
variance $\sigma \propto N^{1/2}$.  The median is shown by the dashed
curve and is much smaller than the mean for small clusters, and slowly
converges to the mean as $N \to \infty$. }  
\label{fig:lfuv}
\end{figure}

To specify the FUV luminosity as a function of stellar mass, $L_{FUV}
(m)$, we use the models of Maeder and collaborators (see Maeder \&
Meynet 1987; Schaller et al. 1992). Specifically, these papers provide
a grid of stellar models as a function of both mass and age. We use
the zero age models to specify the stellar luminosity and effective
temperature. The FUV radiation is dominated by the largest stars,
which reach the main-sequence rapidly. Since the model grids do not
extend to arbitrarily high masses, we enforce a cutoff of 100
$M_\odot$.  Using a blackbody form for the atmosphere, we then
calculate the fraction of the luminosity that is emitted in the FUV
regime (6 eV $< h \nu <$ 13.6 eV). The result is shown in Figure
\ref{fig:lfuv}.

The expectation value $\lexp$ of the FUV luminosity is thus given 
by the integral 
\be
\lexp = \int_1^\infty L_{FUV}(m) {dN \over dm} dm \approx 
8.20 \times 10^{35} \, {\rm erg/s} \, . 
\label{eq:meanfuv} 
\ee 
This expectation value is normalized so that it is the expected FUV
luminosity {\it per star}. Due to the wide range of possible stellar
masses and the sensitive dependence of FUV emission on stellar mass,
this expectation value is much larger than the FUV radiation emitted
by the majority of stars. We thus only expect the FUV radiation from a
given cluster to converge to that implied by the expectation value in the
limit of larger numbers $N$ of stellar members, where large $N$ will
be determined below. Small clusters will often experience large departures 
from the expectation value. 

We want to calculate both the expectation value and its variance for
the FUV luminosities of the entire cluster. The sum given by equation
(\ref{eq:luvsum}) is thus the sum of random variables, where the
variables (individual contributions to the FUV power) are drawn from a
known distribution (determined by the stellar IMF and the $L-m$
relation).  In the limit of large $N$, the expectation value of the
FUV power is thus given by 
\be L_{FUV} (N) = N \lexp \, .
\label{eq:lnexp} 
\ee 
Furthermore, the distribution of possible values for $L_{FUV} (N)$
must approach a gaussian form as $N \to \infty$ because of the central
limit theorem (e.g., Richtmyer 1978), although convergence is slow.  
This gaussian form is (as usual) independent of the form of the initial 
distributions, i.e., it is independent of the stellar IMF and the 
mass-luminosity relation. The width of the distribution also converges 
to a known value and is given by 
\be 
\sigbar^2 = {1 \over N}
\sum_{j=1}^N \sigma_j^2 \quad \Rightarrow \quad \sigbar = \sqrt{N}
\sigma_0 \, , 
\label{eq:nvariance} 
\ee 
where $\sigma_0$ is the width of the individual distribution, i.e., 
\be 
\sigma_0^2 \equiv \langle L_{FUV}^2 \rangle - \lexp^2 \, .  
\ee 
For our usual choice of stellar properties, the dimensionless width
${\widetilde \sigma}_0 = \sigma_0/\lexp \approx 26.4$, i.e., the
variation in the possible values for the FUV luminosity is much
greater than the expectation value. The effective ``signal to noise
ratio'' $S/N$ for variations in FUV luminosity is thus given by
\be 
S/N \equiv {\sigbar \over L_{FUV}(N)} = 
{\sigma_0 \over \sqrt{N} \lexp} \approx {26.4 \over \sqrt{N}} \, .  
\ee 
This definition thus defines a critical value of cluster members
$N_C$, i.e., the number required for the variations in FUV luminosity
to become sufficiently well-defined so that the variations are smaller
than the expectation value. This critical value of $N_C \equiv
{\widetilde \sigma}^2 \sim$ 700.  For $N < N_C$, the FUV radiation
experienced by a solar system living in the cluster is essentially
determined by the largest star in the aggregeate; this largest member
is, in turn, drawn from the probability distribution implied by the
stellar IMF. For clusters with $N > N_C$, it makes sense to consider
expectation values for the FUV radiation provided by the aggregate. 
However, the boundary is not sharp.  Even for $N > N_C$, the total FUV
luminosity $L_{FUV}(N)$ will be subject to substantial fluctuations
from system to system.

We have performed a set of numerical sampling experiments to show that
the mean and variance of the distribution agree with the analytic
predictions derived above. The results are in good agreement. The mean
and variance are shown as a function of cluster size $N$ in Figure
\ref{fig:lfuv}. These numerical experiments show that the convergence
to a purely gaussian distribution is rather slow. The central value
and variance of the actual distribution approach the values predicted
by the central limit theorem much faster than the distribution itself
approaches a normal form. As a result, the median of the distribution
can be significantly different from the mean or expectation value,
especially for cluster with small stellar membership $N$. Figure
\ref{fig:lfuv} also shows the median expected FUV flux as a function
of cluster size $N$. The median is only about 8\% of the mean for
small clusters with $N \approx 100$ and approaches 80\% of the mean
for larger systems ($N = 1000 - 2000$). We note that the median value
provides a better description of the ``typical'' FUV luminosity
produced by a cluster of size $N$. However, the distribution of
possibilities is wide and caution must be taken in characterizing the
probability distribution by only one number.


\begin{figure}
\figurenum{7}
{\centerline{\epsscale{0.90} \plotone{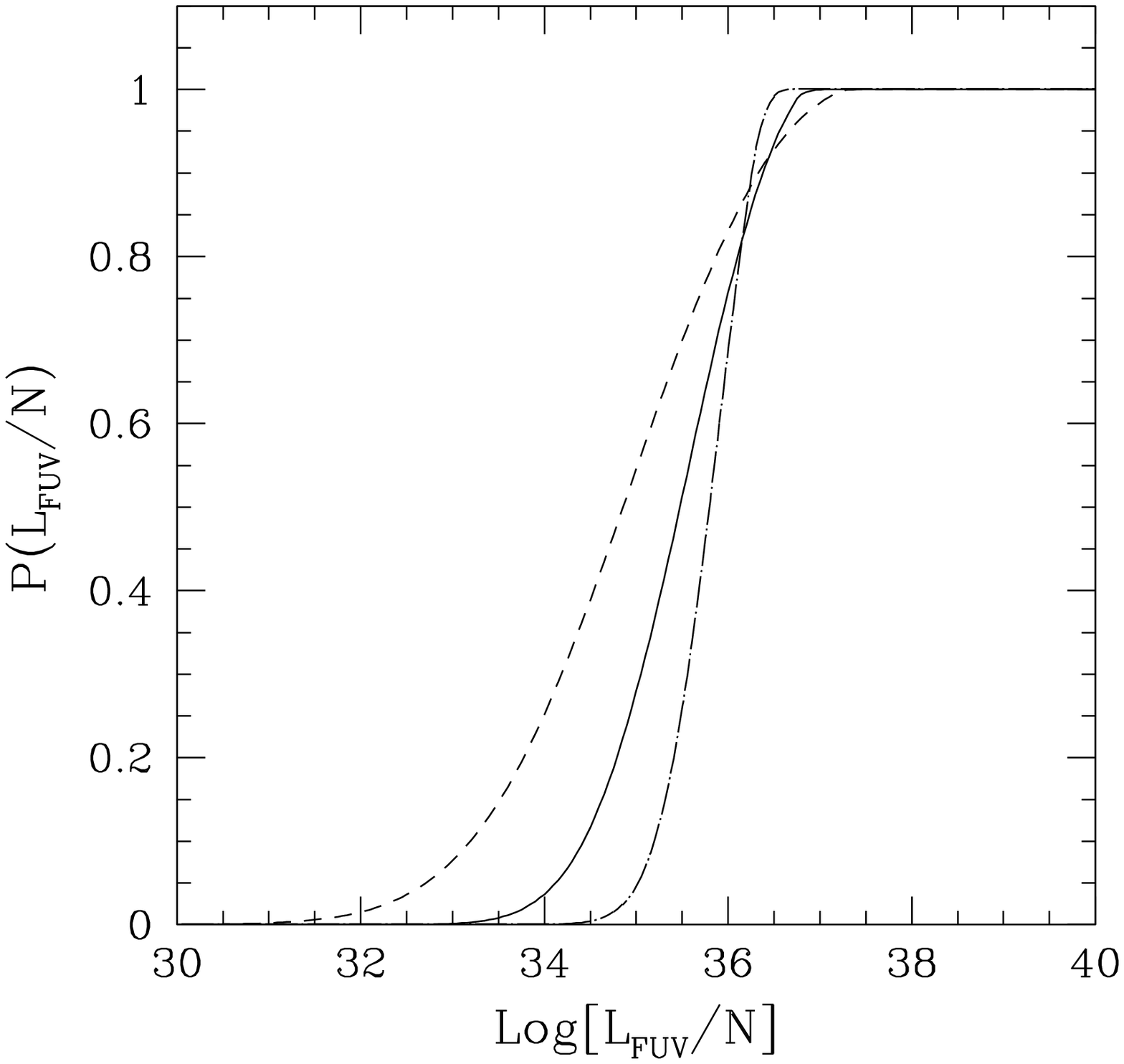}  }}
\figcaption{Cumulative probability distribution for the number of
clusters that produce a given normalized FUV luminosity, where the
normalized FUV luminosity is the total FUV luminosity of the cluster
divided by the number $N$ of cluster members.  The three curves shown
correspond to clusters of size $N$ = 100 (dashed curve), $N$ = 300
(solid curve), and $N$ = 1000 (dot-dashed curve). }
\label{fig:lfuvdist} 
\end{figure}

\begin{figure}
\figurenum{8}
{\centerline{\epsscale{0.90} \plotone{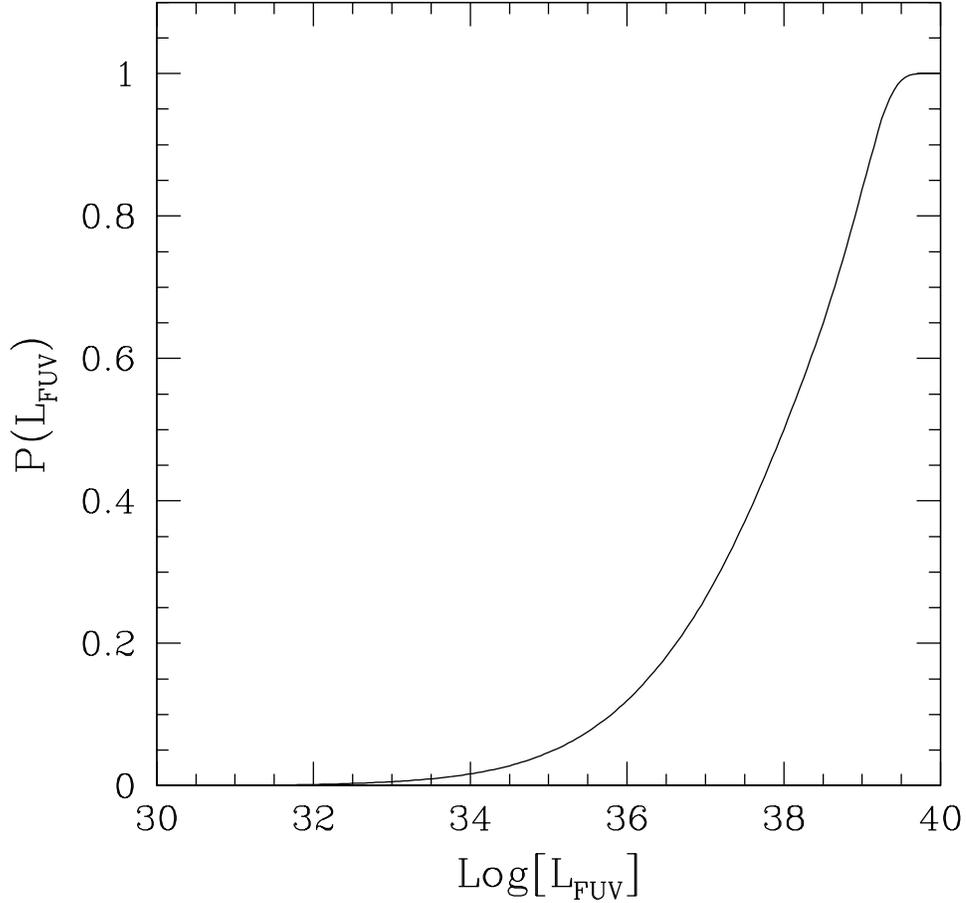}  }}
\figcaption{Cumulative probability distribution for the number of stars 
that live in a cluster with total FUV luminosity $L_{FUV}$ as a function of 
$L_{FUV}$. This distribution is calculated under the assumption that the 
cluster size distribution (the number of stars $N$) follows the data set of 
Lada \& Lada (2003). The size distribution of the data set is sampled 
10,000 times to produce the probability distribution shown here. The median 
FUV luminosity occurs at $L_{FUV} = 10^{38}$ erg/s. Notice that the 
distribution has a long tail at low FUV luminosities. } 
\label{fig:ldistalln} 
\end{figure}

Figure \ref{fig:lfuvdist} shows the probability distributions for FUV
luminosity for three choices of stellar size, namely $N$ = 100, 300,
and 1000. The plot shows the cumulative probability distributions for
the normalized FUV luminosity of the clusters, i.e., the total FUV
luminosity divided by the number of cluster members $N$. In this
representation, the distribution with $N$ = 100 (dashed curve) is the
widest, and the distributions grow narrower with increasing $N$.  In
the limiting case $N \to \infty$, the distribution becomes a step
function at the mean value $\lexp$ (where $\log_{10} \lexp \approx
35.9$). For the three cluster sizes shown here, the normalized
probability distributions converge near $P \approx 0.85$ and
$\log[L_{FUV}/N] \approx 36.2$. This convergence defines a benchmark
for the high end of the distribution, namely, 15\% of the clusters are
expected to have FUV luminosity greater than the limit $L_{FUV} \ge N
(1.6 \times 10^{36})$ erg/s $\approx 2 N \lexp$,

The distributions described above apply to clusters of a given size
$N$ and show how the results depend on $N$. However, since stars are
born in groups/clusters with a distribution of sizes $N$ (Figure
\ref{fig:probn}, Lada \& Lada 2003, Porras et al. 2003), we also need
to determine the overall distribution of FUV luminosities that affects
the entire ensemble of forming solar systems. Toward that end, we
assume that the distribution of stellar groups/clusters is the same as
that of the Lada \& Lada (2003) sample (which is equivalent to that of
Porras et al. 2003). We then sample the distribution of cluster
systems to find a system size $N$, and for each such system we sample
the IMF $N$ times to find the FUV luminosity. Using 10,000
realizations of the cluster sample (corresponding to a total of 127
million stars), we find the cumulative probability distribution for
the expected FUV luminosity. The result is shown in Figure
\ref{fig:ldistalln}. This calculation shows that the median FUV
luminosity experienced by a forming solar system is $10^{38}$ erg/s.
This benchmark cluster luminosity is 122 times the mean FUV luminosity
per star given by equation (\ref{eq:meanfuv}). If every star had the
mean FUV luminosity, this result would imply a typical cluster size $N
\approx 122$; since the typical (median) star has FUV luminosity less
than the mean, the implied typical cluster size is somewhat larger
(consistent with the distribution of Fig. \ref{fig:probn}). Notice
also that individual stellar orbits within the cluster lead to
different radiation exposure -- this issue is discussed below.


\subsection{Effects of FUV Radiation} 

The physical quantity that affects forming solar systems is the FUV
flux, which depends on both the FUV luminosity (\S 3.1) and the radial
position of the solar system within its birth aggregate (\S 2).  For
relatively ``small'' clusters of interest in this paper, we can assume
that the FUV luminosity originates from the few largest stars, which
are generally observed to live near the cluster center (e.g., Testi,
Palla, \& Natta 1999). Here we make the approximation that the FUV
luminosity can be modeled as a point source at the origin. Any given
solar system in orbit within a given cluster will thus experience a
time dependent flux ${\cal F}_{FUV} = L_{FUV} / 4 \pi r^2$, where the
radial position $r$ is specified by the orbit. 

\begin{figure}
\figurenum{9}
{\centerline{\epsscale{0.90} \plotone{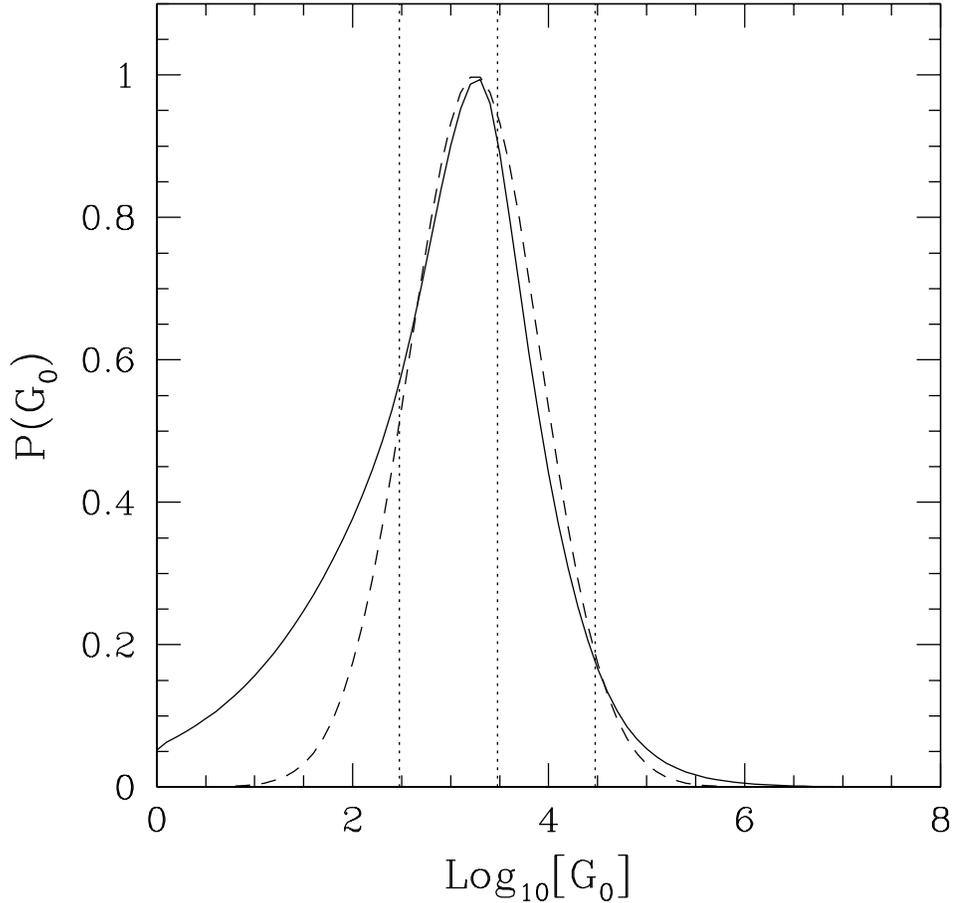}  }}
\figcaption{Probability distribution for FUV flux experienced by the
ensemble of cluster stars as a function of the FUV flux, expressed
here in units of $G_0$ (where $G_0$ = 1 correspondes to a flux of $1.6
\times 10^{-3}$ erg s$^{-1}$ cm$^{-2}$ , which is close to the value
of the interstellar radiation field). The ensemble distribution was
calculated assuming that the number of stars living in groups/clusters
of size $N$ follows the distribution of Figure \ref{fig:probn}, the
radial size $\rstar$ of clusters follows the distribution shown in Figure
\ref{fig:rvsn}, the distribution of FUV luminosity follows that
calculated in \S 3.1, and the density distribution within the cluster
has the simple form $\rho_\ast \propto r^{-1}$ for $0 \le r \le \rstar$.
The dashed curve shows a gaussian distribution with the same peak
location (at $\log_{10} G_0$ = 3.25) and the FWHM (1.575) The calculated
distribution has a significant tail at low flux values.  The vertical
lines at $G_0$ = 300, 3000, and 30000 are benchmark values for which
the effects of photoevaporation on circumstellar disks has been
calculated in detail (Adams et al. 2004). } 
\label{fig:fluxdist} 
\end{figure}

Each cluster provides a distribution of FUV fluxes to its cluster
members, and each ensemble of clusters with a given size $N$ provides
a wider distribution of FUV fluxes to the ensemble of members. Since
clusters come in a distribution of sizes $N$, the collection of all
forming solar systems is thus exposed to a wide distribution of FUV
fluxes.  Figure \ref{fig:fluxdist} shows an estimate for the
probability distribution of FUV flux experienced by the entire
ensemble of cluster stars as a function of flux. In this context, we
express FUV flux in units of $G_0$, where $G_0$ = 1 corresponds to a
benchmark value of $1.6 \times 10^{-3}$ erg s$^{-1}$ cm$^{-2}$ at FUV
wavelengths (close to the value of the interstellar radiation field).
This ensemble distribution for the FUV flux was calculated assuming
that the number of stars living in groups/clusters of size $N$ follows
the distribution shown in Figure \ref{fig:probn} (Lada \& Lada 2003;
Porras et al.  2004), the radial size $\rstar$ of clusters follows the
scaling relation shown in Figure \ref{fig:rvsn}, the distribution of
FUV luminosity follows that calculated in \S 3.1, and the density
distribution within the cluster has the form $\rho_\ast \propto
r^{-1}$ for $0 \le r \le \rstar$. The dashed curve in Figure
\ref{fig:fluxdist} shows a gaussian distribution with the same peak
location and the same FWHM; the true distribution has a substantial
tail at low values and a much smaller tail at high flux values.

In Figure \ref{fig:fluxdist}, the vertical lines at $G_0$ = 300, 3000,
and 30,000 are values for which the photoevaporation of circumstellar
disks due to FUV radiation has been calculated in detail (Adams et
al. 2004). In that study, the value $G_0$ = 3000 was chosen as a
benchmark flux value (which corresponds to a cumulative probability of
$\sim0.74$ for the flux distribution found here). This FUV flux will
drive the evaporation of a circumstellar disk around a 1.0 $M_\odot$
star down to a radius of $r_d \approx 36$ AU in a time of 10 Myr. As a
result, the region of the disk where giant planets form (5 -- 30 AU)
is relatively safe for solar type stars. For smaller stars, however,
flux levels of $G_0$ = 3000 can be significant. A disk orbiting a 0.5
$M_\odot$ (0.25 $M_\odot$) star can be evaporated down to $r_d
\approx$ 18 AU (9 AU) within 10 Myr. As a result, giant planet
formation may be compromised around smaller stars (see also Laughlin
et al. 2004). The results of the detailed photoevaporation models
(Adams et al. 2004) can be summarized by a rough scaling law: An 
FUV flux of $G_0$ = 3000 truncates a circumstellar disk down to a
radius $r_d \approx$ 36 AU $(M_\ast / M_\odot)$, over a time of 10
Myr.


The flux experienced by a ``typical'' star within a cluster can be
characterized in a variety of ways. As shown by Figure
\ref{fig:fluxdist}, however, the distribution of flux values is
extremely wide and cannot be fully described by a single number. If we
consider the composite distribution of Figure \ref{fig:fluxdist} as
representative, then the median FUV flux experienced by a forming
solar system is $G_0$ $\approx$ 900, the peak of the distribution is
at $G_0 \approx$ 1800, and the mean is at $G_0 \approx$ 16,500 (notice
how the mean is much larger than the median).

We can also estimate the typical flux provided by a cluster of a given
size $N$.  This determination requires an integration over both the
distribution of radial positions and the distribution of possible FUV
luminosities. We can use either the mean or medians of these
distributions as the ``typical'' values, but they lead to markedly
different results, as shown in Table 5 below.  For a given FUV
luminosity, the mean radiation flux exposure is given by 
\be 
\langle {\cal F}_{FUV} \rangle = {L_{FUV} \over 4 \pi} \, 
\Bigl\langle {1 \over r^2} \Bigr\rangle \, , 
\ee 
where the average is taken over all stars and over all times. To
evaluate $\langle r^{-2} \rangle$, we use the radial distributions
$N(r)$ calculated from our numerical simulations (analogous to the
mass profiles $M(r)$ given by eq. [\ref{eq:mfit}] and Table 2). 
We can write this expectation value in terms of an effective radius 
$r_{eff}$ defined by 
\be 
r_{eff} \equiv \Bigl\langle {1 \over r^2} \Bigr\rangle^{-1/2} \, .  
\ee 
The effective radii for our six sets of simulations are given in Table
5. Using the mean values of the FUV luminosity in conjunction with
the quantity $\langle r^{-2} \rangle$, we find a ``mean'' flux value,
as listed in the table. Although this value provides one
characterization of the FUV flux, the distributions are extremely
wide, more than an order of magnitude wider than the mean value. As a
result, most of the exposure occurs for those (relatively few) stars
that wander close to the cluster center. The median thus provides a
better measure of the typical FUV flux. The median values for the
radial position, coupled with the median value of the FUV luminosity,
provide an estimate for the ``median'' flux.  The resulting values are
also listed in Table 5. Notice that the flux levels calculated from
the median values of radial position and FUV luminosity are much
smaller than those calculated from the means (by factors of 17 --
190). The mean and median values of the FUV flux, calculated from the
composite distribution of Figure \ref{fig:fluxdist}, are included as
the final line in Table 5.  Notice that the median flux of the
composite ($G_0$ = 900) is smaller than the median values for $N$ =
1000, larger than the values for $N$ = 100, and roughly comparable to
the values for $N$ = 300. This ordering is consistent with finding
that the median (weighted) value of cluster sizes occurs at $N \approx
300$ (from Figure \ref{fig:probn}), i.e., half of the stars in the
sample are found in groups/clusters with $N < 300$ and half are found
in systems with $N > 300$.

\bigskip 
\centerline{\bf Table 5: Expected FUV Flux Values}  

\begin{center} 
\begin{tabular}{lcccc} 
\hline
\hline 
System & $r_{eff}$ & $G_0$ & $r_{med}$ & $G_0$ \\
       & (pc) & -mean- & (pc) & -median-  \\ 
\hline 
100 Cold &   0.080   &  66500   &   0.323   &    359 \\
100 Virial  &   0.112   &  34300   &   0.387   &    250 \\
300 Cold &   0.126   &  81000   &   0.549   &   1550 \\
300 Virial  &   0.181   &  39000   &   0.687   &    992 \\
1000 Cold &  0.197   & 109600   &   0.944   &   3600 \\
1000 Virial  &  0.348   &  35200   &   1.25    &   2060 \\
\hline 
Composite & -- & 16500 & -- & 900 \\  
\hline 
\hline 
\end{tabular} 
\end{center}  

The distributions considered thus far describe the FUV flux
experienced by forming solar systems in a statistical sense.  A
related question is to find the radiation experienced by a given solar
system over the course of its orbit. Here we would like an analytic
description of the orbits in order to see how the results depend on
the relevant physical quantities.  Toward this end, the mean flux 
intercepted by a solar system can be written in the form 
\be 
\langle {\cal F}_{FUV} \rangle = { 1 \over \tau_{1/2} } 
\int {L_{FUV} \over 4 \pi r^2 } \, dt \, , 
\ee
where $\tau_{1/2}$ is the time of a half orbit (e.g., from the inner
turning point to the outer turning point) and the integral is taken
over that same time interval. Here we can model the orbits by assuming
that the total cluster potential has the form of the Hernquist
profile; orbits in this general class of extended mass distributions
have a similar form (for further detail, see Adams \& Bloch 2005,
hereafter AB05). The mean flux can then be written in the form
\be 
\langle {\cal F}_{FUV} \rangle = { \tau_0 \over \tau_{1/2} } 
{L_{FUV} \over 4 \pi r_s^2 } \int_{\xi_1}^{\xi_2}  \, {dt \over \xi^2} 
\, = { \tau_0 \over \tau_{1/2} } {L_{FUV} \over 4 \pi r_s^2 } 
(\Delta \theta) \, , 
\label{eq:meanflux}
\ee
where $\tau_0 \equiv r_s (2 \Psi_0)^{-1/2}$, $\xi = r/r_s$, $r_s$ is
the scale length of the Hernquist profile, $\xi_j$ are the turning
points, and where $\Delta \theta \le \pi$ is the angle subtended by
the half orbit.  The total depth of the gravitational potential well
$\Psi_0$ is given by $\Psi_0 \equiv G M_T/ r_s$, where the mass scale
$M_T = 4 M_1$, where $M_1$ is the mass of the cluster (including both 
stars and gas) contained within the boundary $\rstar$. The orbit time
and turning angle are known functions of the dimensionless energy
$\epsilon \equiv |E|/ \Psi_0$ and angular momentum variable $q \equiv
j^2/2 \Psi_0 r_s^2$ (AB05). To within an accuracy of a few percent,
one can express the dimensionless orbit time and the turning angle by
the functions 
\be 
{\tau_{1/2} \over \tau_0} = \epsilon^{-3/2} 
\bigl( \cos^{-1} \sqrt{\epsilon} + \sqrt{\epsilon} \sqrt{1 - \epsilon} 
\bigr) \, , 
\label{eq:tauorb} 
\ee 
and 
\be 
{\Delta \theta \over \pi} = {1 \over 2} + 
\Bigl[ (1 + 8 \epsilon)^{-1/4} - {1 \over 2} \Bigr]
\Bigl[ 1 + {\ln (q/q_{max}) \over 6 \ln 10 } \Bigr]^{3.6} \, , 
\label{eq:angorb} 
\ee 
where $q_{max}$ represents the maximum angular momentum for a given
energy (that of a circular orbit) and the fitting function is
restricted to the range $10^{-6} \le q/q_{max} \le 1$. The maximum 
value of the angular momentum variable is also a known function of 
dimensionless energy, 
\be 
q_{max} = {1 \over 8 \epsilon} 
{ (1 + \sqrt{1 + 8 \epsilon} - 4 \epsilon )^3 \over 
(1 + \sqrt{1 + 8 \epsilon})^2 } \, . 
\label{eq:qmax} 
\ee 
Equations (\ref{eq:meanflux}) -- (\ref{eq:qmax}) thus specify the
radiation exposure for a solar system on any orbit with given
dimensionless energy $\epsilon$ and angular momentum $q$. The angular
momentum dependence is relatively weak, much weaker than the
dependence on energy. Since $\epsilon > 3/8$ for orbits confined to
the inner part of the Hernquist profile ($\xi < 1$), the turning angle
is confined to the narrow range $(\Delta \theta)/\pi = 1/2 -
\sqrt{2}/2$. We can simplify the flux expression to take the form
\be
\langle {\cal F}_{FUV} \rangle = {L_{FUV} \over 8 r_s^2 } \,
{ A \epsilon^{3/2} \over 
\cos^{-1} \sqrt{\epsilon} + \sqrt{\epsilon} \sqrt{1 - \epsilon} } \, , 
\label{eq:meanflux2} 
\ee
where the parameter $A$ is slowly varying and encapsulates the angular
momentum dependence (with $1 \le A \le \sqrt{2})$.  The leading
coefficient sets the magnitude of the flux. The numerator can be
written as $L_{FUV} = f_N N \lexp$, where $f_N$ is the fraction of the
mean FUV luminosity that the cluster produces. For example, the median
flux corresponds to $f_N$ = 0.088 for $N$ = 100 and $f_N$ = 0.75 for
$N$ = 1000 (see Fig. \ref{fig:lfuv}), whereas the mean flux
corresponds to $f_N$ = 1 by definition.  In the denominator, the scale
length $r_s \sim \rstar$, so that $r_s^2 \sim$ (1 pc)$^2$ $(N/300)$. 
The leading coefficient can thus be evaluated and written in terms of
the interstellar FUV radiation field, i.e., $G \approx 2000 f_N$. The
remaining dimensionless function in equation (\ref{eq:meanflux}) is of
order unity and accounts for the orbital shape.  Deep orbits (close to
the central part of the potential well) with $\epsilon \gta 0.93$ have
values of the dimensionless function greater than unity, whereas lower
energy orbits have values less than unity. In any case, we find $G
\sim 1000$, in agreement with the median values calculated above
(Table 5).

The dynamics of the cluster determine the distributions of energy and
angular momentum parameters ($\epsilon, q$), and these distributions
can be used in conjunction with equation (\ref{eq:meanflux2}) to
calculate the distribution of fluxes experienced by young solar
systems.  For example, for an isotropic velocity distribution and a
density profile $\rho \sim 1/r$, the differential energy distribution
(see BT87) has the form $h(\epsilon) = dm/d\epsilon \propto (1 -
\epsilon)$. Notice, however, that the flux distribution calculated
from the distribution of energy $\epsilon$ (and $q$) is narrower than
that calculated from the distribution of positions (that shown in
Fig. \ref{fig:fluxdist}). The fluxes in the former case are already
averaged over the orbits, whereas those in the latter case are not and
therefore explore a greater range of values.

\section{Scattering Interactions and the Disruption of Planetary Systems} 

The ultimate goal of this section is to calculate the probability that
a solar system will be disrupted as a result of being born within a
group/cluster. In this case, disruption occurs through scattering
interactions with other cluster members, and should thus depend on the
size $N$ of the birth aggregate. In general, the rate of disruption for 
a solar system can be written in the form 
\be 
\Gamma = n \sigma v \, , 
\ee
where $\sigma$ is the disruption cross section, $n$ is the mean
density of other systems, and $v$ is the relative velocity (typically,
$v \sim 1$ km/s).  In this case, however, the results of $N$-body
simulations provide the rate at which solar system experience close
encounters within a closest approach distance $b$ as a function of
$b$. As a result, we only need to determine how close such scattering
encounters must be in order for disruption to take place. Here we make
the approximation that the cluster dynamics (which determines the rate
of close encounters) is independent of the scattering dynamics between
a solar system and a passing star (which determines the cross
sections).

Using our planet scattering code developed previously (Laughlin \&
Adams 2000; Adams \& Laughlin 2001), we can calculate the cross
sections for the disruption of solar systems with varying stellar
masses (see also Heller 1993, 1995; de la Fuenta Marcos \& de la
Fuente Marcos 1997, 1999).  Since most stars passing nearby with the
potential for disrupting a solar system are binaries, we want to find
the effective cross section $\cross$ for a specified change in orbital
parameters resulting from scattering encounters with binaries. We
define this effective cross section $\cross$ through the relation
\be
\cross \equiv \int_{0}^{\infty} f_D (a) (4 \pi a^{2}) p(a) \, da \, , 
\label{eq:netcross} 
\ee
where $a$ is the semimajor axis of the binary orbit and $p(a)$  
specifies the probability of encountering a binary system with a given
value of $a$.  Notice that for a given value of $a$, this integal only
includes those scattering interactions that fall within the
predetermined area $4 \pi a^{2}$, where the numerical coefficient (4)
is chosen to be large enough to include all relevant interactions and
small enough to allow for finite computing resources (in principle,
one should include all interactions, no matter how distant).  In
practice, we find that the area $4 \pi a^2$ provides a good compromise
between accuracy and computational expediency. The function $f_D (a)$
specifies the fraction of the encounters that result in a particular 
outcome (e.g., a given change in the orbital parameters of the solar
system), and the determination of $f_D$ is the main computational task. 
The distribution $p(a)$ is determined by the observed distribution 
of binary periods and by the normalization condition 
\be 
\int_{0}^{\infty} p(a) da = 1 \, , 
\ee 
where we use observational results to specify the period distribution
(Kroupa 1995a).  The observed distribution is extremely broad, with
roughly equal numbers of systems in each logarithmic interval in
semimajor axis $a$ and with a broad overall peak falling near $P$ =
$10^5$ days.  Although the distribution includes binaries with periods
longer than $10^{7}$ days, this set of scattering experiments only
includes binaries with $a < 1000$ AU because wider binaries do not
contribute appreciably to the cross sections.
 
The set of possible encounters that can occur between a given solar
system and a field binary is described by 10 basic input parameters.
These variables include the binary semimajor axis $a$, the stellar
masses, $m_{\ast 1}$ and $m_{\ast 2}$, of the binary pair, the
eccentricity $e_{\rm b}$ and the initial phase angle $\ell_{\rm b}$ of
the binary orbit, the asymptotic incoming velocity $v_{\rm inf}$ of
the solar system with respect to the center of mass of the binary, the
angles $\theta$, $\psi$, and $\phi$ which describe the impact
direction and orientation, and finally the impact parameter $h$ of the
collision.
Additional (intrinsic) parameters are required to specify the angular
momentum vector and initial orbital phases of the planets within the
solar system. In this study, we consider a class of solar systems in
which the central stellar mass varies, but the planetary orbits are
always taken to be those of our solar system.  In other words, each
solar system has an analog of Jupiter, a planet with one Jupiter mass
$m_J$ in an initial orbit of semimajor axis $a_J$ = 5 AU; similarly,
each solar system has an analog of Saturn, Uranus, and Neptune.  The
planetary orbits are assumed to be circular (initially) so that we can
determine the change in orbital parameters from a known baseline. All
of the planetary properties are kept fixed so we can isolate the
effects of changing the mass of the central star.

In order to compute the fraction of disruptive encounters $f_D (a)$
and the corresponding cross sections, we perform a large number of
scattering experiments using a Monte Carlo scheme to select the input
parameters.  This section reports on the results from more than $10^5$
such scattering experiments.  Individual encounters are treated as
$N$-body problems in which the equations of motion are integrated
using a Bulirsch-Stoer scheme (Press et al. 1986).  During each
encounter, we require that overall energy conservation be maintained
to an accuracy of at least one part in $10^{8}$. For most experiments,
energy and angular momentum are conserved to one part in $10^{10}$.
This high level of accuracy is needed because we are interested in the
resulting planetary orbits, which carry only a small fraction of the
total angular momentum and orbital energy of the system.
 
For each scattering experiment, the initial conditions are drawn from
the appropriate parameter distributions. More specifically, the binary
eccentricities are sampled from the observed distribution (Duquennoy
\& Mayor 1991).  The masses of the two binary components are drawn
separately from a log-normal initial mass function (IMF) that is
consistent with the observed distribution of stellar masses (for
completeness, we note that the IMF has a power-law tail at high
masses, although this departure will not affect the cross sections
calculated here).  For both the primary and the secondary, we enforce
a lower mass limit of 0.075 $M_\odot$ and hence our computed
scattering results do not include brown dwarfs.  This cutoff has a
relatively small effect because our assumed IMF peaks in the stellar
regime. The impact velocities at infinite separation, $v_{\rm inf}$,
are sampled from a Maxwellian distribution with dispersion
$\sigma_{v}$ = 1 km/s, which is a typical value for stellar clusters
with the range of parameters considered here (BT87).  The initial
impact parameters $h$ are chosen randomly within a circle of radius
$2a$ centered on the binary center of mass (a circular target area of
$4 \pi a^2$ as in eq.  [\ref{eq:netcross}]).
 
Using the Monte Carlo technique outlined above, we have performed
$N_{\rm exp}$ $\approx$ 20,000 scattering experiments for collisions
between binary star systems and solar systems of each stellar mass
(i.e., a total of $\sim10^5$ simulations).  We use logarithmically
spaced mass values: $M_\ast/M_\odot$ = 2, 1, 1/2, 1/4, and 1/8.  As
described above, all solar systems are taken to have the architecture
of our outer solar system.  From the results of these 7-body
scattering experiments, we compute the cross sections for orbital
disruption of each outer planet (according to equation
[\ref{eq:netcross}]). Note that the procedure described thus far
implicitly asssumes that all passing stars are binary. Although the
majority of solar type stars have binary companions, smaller stars
(e.g., M stars) have a lower binary fraction. The true cross sections
should thus be written as $\cross_T = F_b \cross_{\rm b} + (1 - F_b)
\cross_{\rm s}$, where $F_b$ is the binary fraction and $\cross_{\rm
s}$ is the cross section of interactions between single stars and
solar systems (and is not calculated here). However, $\cross_{\rm s}
\ll \cross_{\rm b}$ (Adams \& Laughlin 2001) so that to a good working 
approximation one can use $\cross_T \approx F_b \cross_{\rm b}$.

\begin{figure}
\figurenum{10}
{\centerline{\epsscale{0.90} \plotone{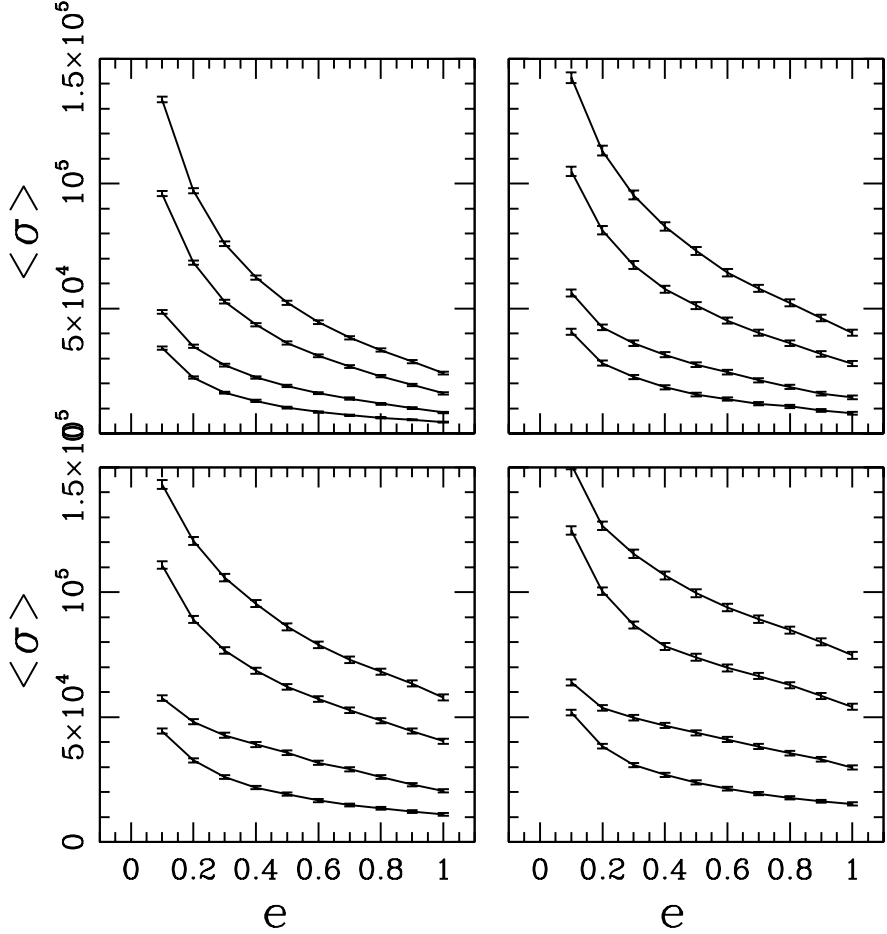} }} 
\figcaption{Scattering cross sections for solar systems to increase
the eccentricity $e$ of planetary orbits, plotted here as a function of
eccenctricity. All cross sections are given in units of (AU)$^2$.  The
four panels shown here correspond to the four largest stellar mass
values of our computational survey, i.e., $M_\ast$ = 2.0 $M_\odot$
(upper left) 1.0 $M_\odot$ (upper right), 0.5 $M_\odot$ (lower left),
and 0.25 $M_\odot$ (lower right). In each panel, the four curves shown
correspond to four giant planets orbiting the central star, where the
planets have the same masses and starting semimajor axes as the giant
planets in our solar system. The top curve in each panel corresponds
to an analog of Neptune and the bottom curve corresponds to an analog
of Jupiter. The cross section for increasing the eccentricity beyond
unity (right end points of the curves) corresponds to ejection of the
planet. The error bars correspond to the uncertainties incurred due to
the (incomplete) Monte Carlo sampling of the parameter space. }
\label{fig:crossfour} 
\end{figure}

The cross sections for the planets to increase their orbital
eccentricities are shown in Tables 6 -- 10 for varying stellar mass
(see also Fig. \ref{fig:crossfour}).  For each given value of
eccentricity $e$, the table entries give the cross sections [in units
of (AU)$^2$] for the eccentricity to increase to any value greater
than the given $e$; these cross sections include events leading to
either ejection of the planet or capture by another star. The listings
for $e$ = 1 thus give the total cross sections for planetary escape
and capture (taken together). The last two lines of the Tables present
the cross sections for planetary escapes and captures separately.  For
each cross section listed, the Tables also provide the one standard
deviation error estimate [also in (AU)$^2$] for the Monte Carlo
integral; this quantity provides a rough indication of the errors due
to the statistical sampling process (Press et al. 1986).  Figure
\ref{fig:crossfour} shows the cross sections plotted as a function of
the eccentricity increase for each of the four planets and for the
four largest stellar mass values.


\newpage 
\centerline{\bf Table 6: Scattering Cross Sections for $M_\ast = 2.0 M_\odot$ Stars} 
\begin{center} 
\begin{tabular}{lcccc} 
\hline
\hline 
$e$/outcome & $\cross$ (Jupiter) & $\cross$ (Saturn) & $\cross$ (Uranus) & $\cross$ (Neptune) \\ 
\hline  
 0.10 &   34200  $\pm$  632 &   48600  $\pm$  738 &   96000  $\pm$ 1030 &  133700  $\pm$ 1180 \\
 0.20 &   22300  $\pm$  513 &   34900  $\pm$  631 &   68400  $\pm$  881 &   97200  $\pm$ 1040 \\
 0.30 &   16300  $\pm$  436 &   27300  $\pm$  563 &   52800  $\pm$  782 &   75900  $\pm$  926 \\
 0.40 &   13000  $\pm$  389 &   22400  $\pm$  511 &   43500  $\pm$  714 &   62400  $\pm$  846 \\
 0.50 &   10300  $\pm$  343 &   18900  $\pm$  471 &   36100  $\pm$  653 &   52300  $\pm$  780 \\
 0.60 &    8560  $\pm$  314 &   16100  $\pm$  439 &   31100  $\pm$  609 &   44600  $\pm$  725 \\
 0.70 &    7300  $\pm$  291 &   13900  $\pm$  410 &   26800  $\pm$  569 &   38300  $\pm$  677 \\
 0.80 &    6240  $\pm$  269 &   11900  $\pm$  380 &   22900  $\pm$  526 &   33300  $\pm$  635 \\
 0.90 &    5470  $\pm$  253 &   10100  $\pm$  351 &   19400  $\pm$  486 &   28700  $\pm$  595 \\
 1.00 &    4550  $\pm$  231 &    8350  $\pm$  321 &   16000  $\pm$  446 &   24200  $\pm$  555 \\
 escape  &    4010  $\pm$  216 &    7590  $\pm$  306 &   13900  $\pm$  413 &   20600  $\pm$  504 \\
 capture &     539  $\pm$   83 &     764  $\pm$   99 &    2100  $\pm$  168 &    3640  $\pm$  232 \\
\hline 
\hline 
\end{tabular} 
\end{center}  

\bigskip 
\centerline{\bf Table 7: Scattering Cross Sections for $M_\ast = 1.0 M_\odot$ Stars} 
\begin{center} 
\begin{tabular}{lcccc} 
\hline
\hline 
$e$/outcome & $\cross$ (Jupiter) & $\cross$ (Saturn) & $\cross$ (Uranus) & $\cross$ (Neptune) \\ 
\hline  
 0.10 &   40700  $\pm$ 1190 &   56300  $\pm$ 1380 &  104900  $\pm$ 1860 &  142400  $\pm$ 2110 \\
 0.20 &   28100  $\pm$  996 &   42500  $\pm$ 1200 &   81300  $\pm$ 1650 &  113200  $\pm$ 1910 \\
 0.30 &   22600  $\pm$  895 &   36100  $\pm$ 1110 &   67400  $\pm$ 1510 &   95400  $\pm$ 1780 \\
 0.40 &   18500  $\pm$  801 &   31500  $\pm$ 1040 &   57700  $\pm$ 1400 &   82900  $\pm$ 1670 \\
 0.50 &   15500  $\pm$  731 &   27500  $\pm$  974 &   51200  $\pm$ 1330 &   73100  $\pm$ 1580 \\
 0.60 &   13700  $\pm$  692 &   24500  $\pm$  924 &   45200  $\pm$ 1250 &   64300  $\pm$ 1480 \\
 0.70 &   11900  $\pm$  640 &   21300  $\pm$  865 &   40300  $\pm$ 1190 &   58000  $\pm$ 1420 \\
 0.80 &   10800  $\pm$  606 &   18600  $\pm$  812 &   36100  $\pm$ 1130 &   52200  $\pm$ 1360 \\
 0.90 &    9270  $\pm$  564 &   16000  $\pm$  754 &   31800  $\pm$ 1060 &   46200  $\pm$ 1280 \\
 1.00 &    8040  $\pm$  529 &   14400  $\pm$  728 &   28000  $\pm$ 1010 &   40300  $\pm$ 1220 \\
 escape  &    7060  $\pm$  488 &   13100  $\pm$  688 &   24900  $\pm$  952 &   33900  $\pm$ 1100 \\
 capture &     973  $\pm$  203 &    1350  $\pm$  238 &    3010  $\pm$  347 &    6390  $\pm$  517 \\
\hline 
\hline 
\end{tabular} 
\end{center}  

\newpage 
\centerline{\bf Table 8: Scattering Cross Sections for $M_\ast = 0.50 M_\odot$ Stars} 
\begin{center} 
\begin{tabular}{lcccc} 
\hline
\hline 
$e$/outcome & $\cross$ (Jupiter) & $\cross$ (Saturn) & $\cross$ (Uranus) & $\cross$ (Neptune) \\ 
\hline  
 0.10 &   44400  $\pm$ 1010 &   57700  $\pm$ 1140 &  110900  $\pm$ 1560 &  143200  $\pm$ 1730 \\
 0.20 &   32700  $\pm$  870 &   48200  $\pm$ 1050 &   89100  $\pm$ 1400 &  120600  $\pm$ 1610 \\
 0.30 &   26000  $\pm$  775 &   42800  $\pm$  993 &   76600  $\pm$ 1310 &  105900  $\pm$ 1520 \\
 0.40 &   21800  $\pm$  708 &   39100  $\pm$  952 &   68600  $\pm$ 1250 &   95500  $\pm$ 1460 \\
 0.50 &   19200  $\pm$  664 &   35800  $\pm$  914 &   62200  $\pm$ 1190 &   86200  $\pm$ 1390 \\
 0.60 &   16600  $\pm$  618 &   31700  $\pm$  860 &   57300  $\pm$ 1150 &   78900  $\pm$ 1340 \\
 0.70 &   14800  $\pm$  585 &   29200  $\pm$  828 &   52900  $\pm$ 1110 &   72900  $\pm$ 1290 \\
 0.80 &   13500  $\pm$  557 &   26000  $\pm$  782 &   48600  $\pm$ 1070 &   68200  $\pm$ 1260 \\
 0.90 &   12200  $\pm$  531 &   23000  $\pm$  732 &   44400  $\pm$ 1020 &   63500  $\pm$ 1220 \\
 1.00 &   11100  $\pm$  510 &   20500  $\pm$  693 &   40400  $\pm$  982 &   57900  $\pm$ 1190 \\
 escape  &   10400  $\pm$  496 &   18900  $\pm$  665 &   36200  $\pm$  922 &   50600  $\pm$ 1090 \\
 capture &     701  $\pm$  119 &    1560  $\pm$  194 &    4150  $\pm$  338 &    7300  $\pm$  458 \\
\hline 
\hline 
\end{tabular} 
\end{center}  

\bigskip 
\centerline{\bf Table 9: Scattering Cross Sections for $M_\ast = 0.25 M_\odot$ Stars} 
\begin{center} 
\begin{tabular}{lcccc} 
\hline
\hline 
$e$/outcome & $\cross$ (Jupiter) & $\cross$ (Saturn) & $\cross$ (Uranus) & $\cross$ (Neptune) \\ 
\hline  
 0.10 &   51900  $\pm$ 1140 &   63900  $\pm$ 1250 &  124800  $\pm$ 1720 &  151100  $\pm$ 1850 \\
 0.20 &   38400  $\pm$  979 &   53800  $\pm$ 1150 &  100500  $\pm$ 1560 &  126700  $\pm$ 1710 \\
 0.30 &   30900  $\pm$  875 &   49700  $\pm$ 1110 &   86900  $\pm$ 1460 &  115400  $\pm$ 1650 \\
 0.40 &   26900  $\pm$  810 &   46700  $\pm$ 1080 &   78300  $\pm$ 1390 &  106800  $\pm$ 1600 \\
 0.50 &   23900  $\pm$  761 &   43800  $\pm$ 1050 &   74000  $\pm$ 1350 &   99700  $\pm$ 1550 \\
 0.60 &   21300  $\pm$  717 &   41100  $\pm$ 1020 &   69700  $\pm$ 1320 &   93900  $\pm$ 1510 \\
 0.70 &   19400  $\pm$  683 &   38300  $\pm$  988 &   66500  $\pm$ 1290 &   89300  $\pm$ 1480 \\
 0.80 &   17700  $\pm$  655 &   35600  $\pm$  959 &   62800  $\pm$ 1260 &   84900  $\pm$ 1450 \\
 0.90 &   16300  $\pm$  630 &   33200  $\pm$  929 &   58600  $\pm$ 1220 &   80200  $\pm$ 1410 \\
 1.00 &   15300  $\pm$  614 &   29900  $\pm$  880 &   54100  $\pm$ 1190 &   74800  $\pm$ 1400 \\
 escape  &   14500  $\pm$  597 &   27800  $\pm$  842 &   49300  $\pm$ 1120 &   65300  $\pm$ 1290 \\
 capture &     741  $\pm$  140 &    2040  $\pm$  257 &    4880  $\pm$  394 &    9440  $\pm$  546 \\
\hline 
\hline 
\end{tabular} 
\end{center}  

\newpage 
\centerline{\bf Table 10: Scattering Cross Sections for $M_\ast = 0.125 M_\odot$ Stars} 
\begin{center} 
\begin{tabular}{lcccc} 
\hline
\hline 
$e$/outcome & $\cross$ (Jupiter) & $\cross$ (Saturn) & $\cross$ (Uranus) & $\cross$ (Neptune) \\ 
\hline  
 0.10 &   65100  $\pm$ 1170 &   77700  $\pm$ 1260 &  147500  $\pm$ 1690 &  208200  $\pm$ 1940 \\
 0.20 &   50200  $\pm$ 1030 &   67800  $\pm$ 1190 &  111600  $\pm$ 1480 &  134700  $\pm$ 1590 \\
 0.30 &   37600  $\pm$  888 &   63800  $\pm$ 1160 &   99400  $\pm$ 1410 &  121400  $\pm$ 1530 \\
 0.40 &   30800  $\pm$  796 &   61500  $\pm$ 1140 &   92400  $\pm$ 1370 &  113600  $\pm$ 1490 \\
 0.50 &   26700  $\pm$  734 &   58900  $\pm$ 1120 &   87200  $\pm$ 1340 &  107200  $\pm$ 1450 \\
 0.60 &   24000  $\pm$  694 &   56000  $\pm$ 1100 &   83100  $\pm$ 1310 &  101800  $\pm$ 1420 \\
 0.70 &   21800  $\pm$  664 &   52700  $\pm$ 1070 &   79100  $\pm$ 1280 &   97800  $\pm$ 1400 \\
 0.80 &   20000  $\pm$  637 &   49300  $\pm$ 1040 &   75100  $\pm$ 1250 &   93800  $\pm$ 1380 \\
 0.90 &   18600  $\pm$  612 &   45200  $\pm$ 1000 &   72100  $\pm$ 1230 &   90100  $\pm$ 1350 \\
 1.00 &   17200  $\pm$  588 &   40100  $\pm$  948 &   65900  $\pm$ 1190 &   84700  $\pm$ 1340 \\
 escape  &   16100  $\pm$  564 &   38000  $\pm$  920 &   60100  $\pm$ 1130 &   74500  $\pm$ 1240 \\
 capture &    1060  $\pm$  167 &    2160  $\pm$  231 &    5750  $\pm$  386 &   10200  $\pm$  524 \\
\hline 
\hline 
\end{tabular} 
\end{center}  

\bigskip 
\centerline{\bf Table 11: Scattering Cross Sections for Angular Increase} 
\begin{center} 
\begin{tabular}{cccccc} 
\hline
\hline 
$\Delta i$ & $\cross$ & $\cross$ & $\cross$ & $\cross$ & $\cross$ \\
(degrees) & 0.125 $M_\odot$ & 0.25 $M_\odot$ & 0.5 $M_\odot$ & 1.0 $M_\odot$ & 2.0 $M_\odot$ \\ 
\hline  
10.0   &  108800   &  104400   &   97400   &   89600   &   75500 \\
20.0   &   91100   &   85600   &   78000   &   65800   &   50700 \\
30.0   &   83200   &   77300   &   68000   &   55100   &   40400 \\
40.0   &   77600   &   71900   &   61600   &   48800   &   34400 \\
50.0   &   72800   &   67600   &   55900   &   43700   &   30200 \\
60.0   &   68000   &   62500   &   51900   &   40100   &   26500 \\
70.0   &   63000   &   57400   &   47900   &   36500   &   24000 \\
80.0   &   58300   &   53000   &   43600   &   33700   &   21800 \\
90.0   &   54100   &   48600   &   40300   &   30700   &   19400 \\
\hline 
\hline 
\end{tabular} 
\end{center}  

\begin{figure}
\figurenum{11}
{\centerline{\epsscale{0.90} \plotone{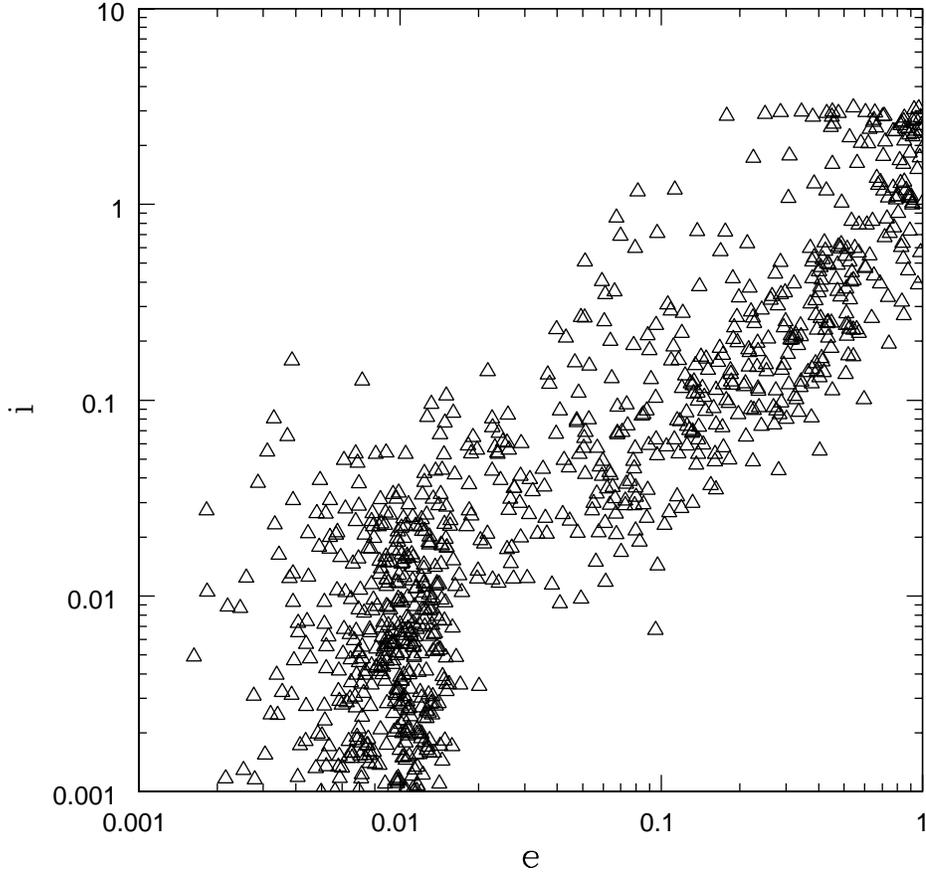} }} 
\figcaption{Scatter plot showing the correlation of changes in
eccentricity with changes in the orbital inclination angle (in radians). 
Each symbol shows the result of one scattering simulation (not all cases 
are shown). The inclination angle is defined to be the maximum angle
between the the perturbed orbital planes of the planets and the
original orbital plane. The eccentricity considered here is the
(final) eccentricity of the Neptune analog. This plot shows results
from the scattering experiments using stellar mass $M_\ast$ = 1.0 
$M_\odot$; the results are similar for all five stellar masses 
considered herein. }
\label{fig:eicor} 
\end{figure}

Another way for solar systems to be disrupted is by changing the
planes of the planetary orbits. One can use the results of the Monte
Carlo scattering calculations to compute the cross sections for the
inclination angles of the planetary orbits to increase by varying
amounts. These results are shown as a function of angle in Table 11
for the five stellar mass values used here.  More specifically, we
define the inclination angle increase $\Delta i$ to be the maximum
angle between the orbital plane of the perturbed (post-encounter)
planets and the original orbital plane.  In Table 11, the Monte Carlo
uncertainties are not listed, but the sampling statistics are good,
and the effective errors are approximately 2 percent. In addition, we
find that the increases in inclination angle are well correlated with
the predicted increases in eccentricity, as shown in Figure
\ref{fig:eicor}.  For the five stellar masses considered here, the
linear correlation coefficient between the inclination angle increase
and the (final) eccentricity of the Neptune-analog planet lies in the
range $\cal R$ = 0.70 -- 0.75.


The cross sections scale roughly with the inverse square root of the
stellar mass. For example, the total ejection cross section is one of
the more useful quantities considered here. We find that the mass
dependence of the cross section for a given planet to be ejected can
be written in the form
\be
\cross_{\rm ej} \, (M_\ast / M_\odot)^{1/2} \approx 
{\cal C}_P \approx constant \, , 
\ee 
where the constant ${\cal C}_P$ depends on which planet is being ejected. 
For Jupiter, Saturn, Uranus, and Neptune, respectively, we find 
${\cal C}_J$ = 7200 $\pm$ 800 AU$^2$, 
${\cal C}_S$ = 14,000 $\pm$ 1100 AU$^2$, 
${\cal C}_U$ = 25,900 $\pm$ 2440 AU$^2$, and 
${\cal C}_N$ = 36,600 $\pm$ 4070 AU$^2$. 
When we scale the cross sections by the mass of the central star, the
scaling law is not perfect, but rather retains some variation that is
quantified by the quoted ``error bars'' given here. Next we note that
these cross sections almost scale linearly with the semimajor axes of
the planet. If we perform such a scaling, the ejection cross section
can be written in the form
\be
\cross_{\rm ej} \approx {\cal C}_0 \, (a_p / {\rm AU}) \, 
(M_\ast / M_\odot)^{-1/2} \, , 
\ee 
where ${\cal C}_0 = 1350 \pm 160$ AU$^2$ and where $a_p$ is the 
semimajor axis of the planetary orbit. 

Now we can put the pieces together and apply these results to
clusters.  The output measures from the numerical simulations show
that the rates of close encounters have the form $\Gamma = \Gamma_0
(r/r_0)^\gamma$, where the parameters $\Gamma_0$ and $\gamma$ depend
on the starting conditions in the cluster. The length scale $r_0$ =
1000 AU defines the units. The rate of ejection of planets is thus
given by
\be 
\Gamma_{eject} = \Gamma_0 \, 
\Bigl[ { {\cal C}_0 (a_p/{\rm AU}) \over \pi r_0^2} \Bigr]^{\gamma/2} 
\, \Bigl( { M_\ast \over M_\odot } \Bigr)^{-\gamma/4}  \, . 
\ee 
This expression gives the ejection rate per star for a given $M_\ast$. 
To find the total ejection rate for the cluster, one must integrate 
over the IMF, normalized to the cluster size $N$, i.e., $\int dm 
(dN/dm) m^{-\gamma/4}$ where $\int (dN/dm) dm$ = $N$.  For example,
the rate of ejection of planetary analogs of Jupiter in a cluster of
$N$ = 300 stars with a cold starting condition can be readily found:
The numerical simulations provide $\Gamma_0$ = 0.096 (interactions per
star per Myr) and $\gamma$ = 1.7 (see Table 3). The ejection rate of
Jupiters is thus $\Gamma_J \approx 0.15$ ejections per cluster per
Myr. Over the 10 Myr lifetime spanned by the simulations, only 1 or 2
Jupiter ejections are expected per cluster (these results are
consistent with those obtained by Smith \& Bonnell 2001 and by de le
Fuente Marcos \& de la Fuente Marcos 1997).  The number of ejected
planets is not only small, but it is much smaller than the number of
ejections expected from internal (planet-planet) scattering events
(Moorhead \& Adams 2005).  Even for the larger cross section for the
ejection of Neptunes, the number of expected events is only about
7. For solar systems orbiting smaller stars, e.g., with mass $M_\ast$
= 0.25 $M_\odot$, the ejection cross sections and hence the expected
number of ejected planets are larger by a factor of $\sim 2$. Of
course, smaller stars may have trouble forming planets due to
increased efficacy of disk evaporation (\S 3, Adams et al. 2004) and
other difficulties (Laughlin et al. 2004).


\begin{figure}
\figurenum{12}
{\centerline{\epsscale{0.90} \plotone{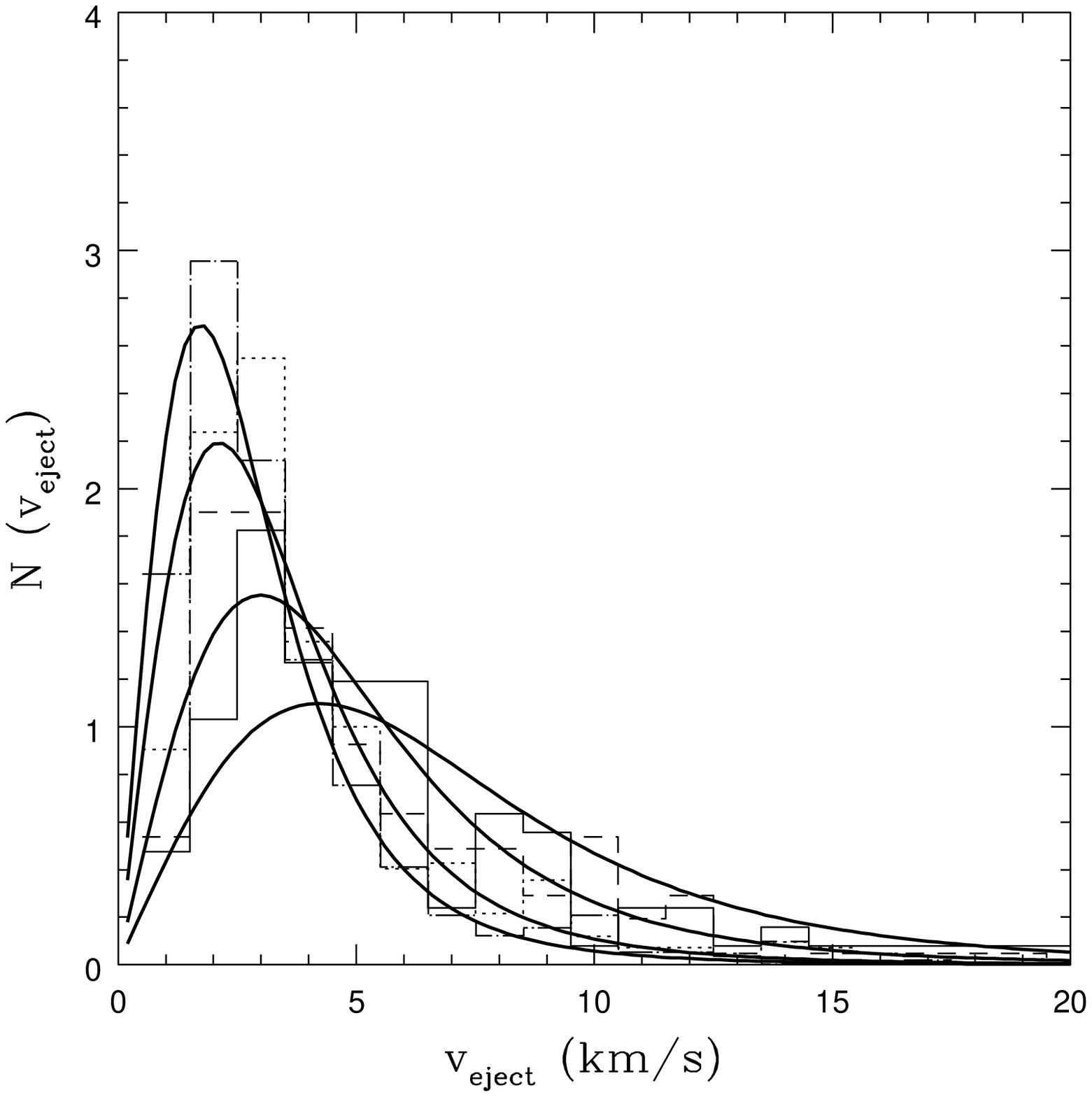} }} 
\figcaption{Distribution of ejection speeds for scattering
interactions with central stellar mass $M_\ast$ = 0.50 $M_\odot$.  
The histograms show the distributions of ejection speeds found in 
the numerical simulations for the analogs of Jupiter (solid), Saturn 
(dashes), Uranus (dots), and Neptune (dot-dashes).  The four smooth
solid curves show the expected distribution from the simple theory
outlined in the text.}
\label{fig:eject50}
\end{figure}

\begin{figure}
\figurenum{13}
{\centerline{\epsscale{0.90} \plotone{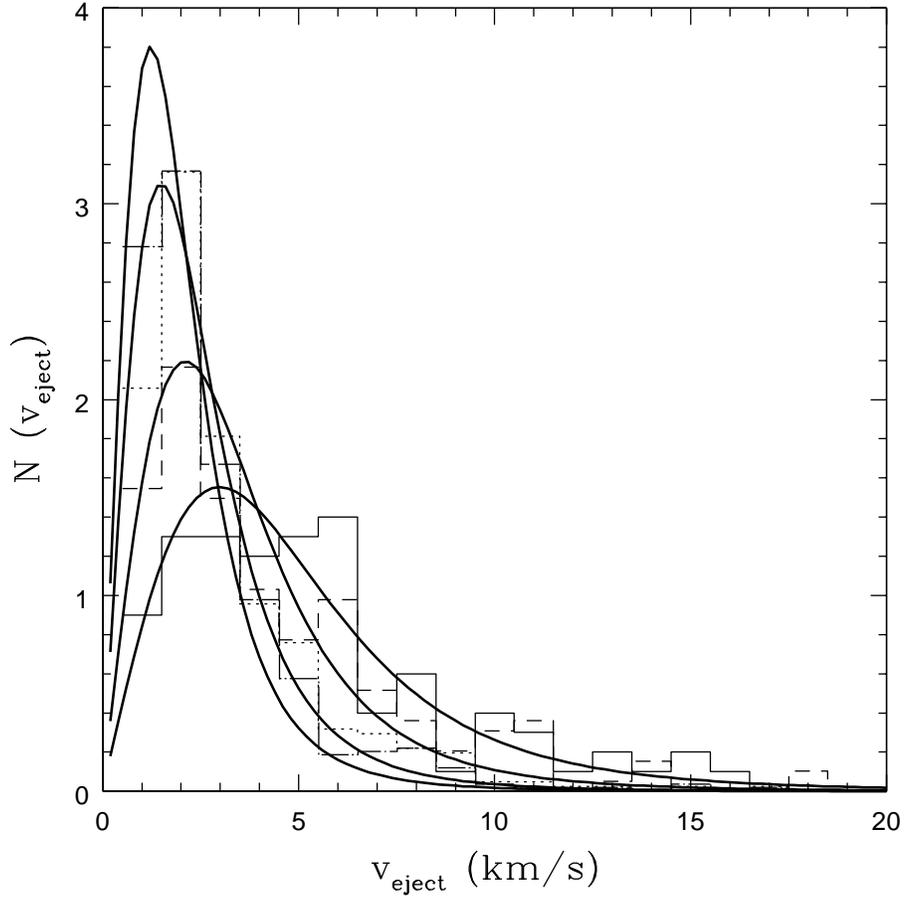} }} 
\figcaption{Distribution of ejection speeds for scattering 
interactions with central stellar mass $M_\ast$ = 0.25 $M_\odot$.  
The histograms show the distributions of ejection speeds found in 
the numerical simulations for the analogs of Jupiter (solid), Saturn 
(dashes), Uranus (dots), and Neptune (dot-dashes).  The four smooth
solid curves show the expected distribution from the simple theory
outlined in the text.}
\label{fig:eject25}
\end{figure}

Another result from our ensemble of scattering experiments is the
distribution of ejection speeds for planets that are removed from
their solar systems during the interaction. The resulting
distributions are shown for each of the four giant planets in Figure
\ref{fig:eject50} (for stellar mass $M_\ast$ = 0.50 $M_\odot$) and
Figure \ref{fig:eject25} ($M_\ast$ = 0.25 $M_\odot$). Also shown are
the theoretically expected distributions based on the idea that
ejections involve sufficiently close encounters that the gravitational
potential of the perturber (the passing star) dominates that of the
central star.  This type of interaction implies a distribution of
ejection speeds of the basic form
\be
{dp \over du} = {4 u \over (1 + u^2)^3} \, , 
\ee 
where $u = v/v_0$ and the velocity scale is given by $v_0^2$ = $G
M_\ast/a$ (Moorhead \& Adams 2005). Notice that the impact speed of
the binary $(v_{\rm inf})$ does not enter into this formula because
$v_0 \gg v_{\rm inf}$.  As shown in Figures \ref{fig:eject50} and
\ref{fig:eject25}, this type of distribution provides a good fit to
that found in the simulation data. Note that the numerical and
analytic distributions are given the same normalization for all four
planets. The overall number of ejections will vary with the planet's
semimajor axes, as given by the cross sections in Tables 6 -- 10.

\section{NGC 1333 -- A Case Study}

The recent identification of 93 N$_2$H$^+$ (1-0) clumps in the young
cluster NGC 1333 by Walsh et al. (2004ab) provides an excellent
opportunity to apply the theoretical program developed in this paper
toward the understanding of an observed cluster.  More specifically,
the data provide two dimensional (2D) position measurements (in the
plane of the sky) and one dimensional (1D) velocity measurements (in
the line of sight) for each of the 93 clumps. As a result, we need to
reconstruct the remaining three phase space variables in order to make
full three dimensional (3D) simulations of the cluster. Because the
reconstruction process contains a random element (see below), we have
to perform multiple realizations of the simulations in order to
describe the dynamics.  In addition, because the data do not
completely specify the starting conditions (without reconstruction),
this set of simulations represents a ``theoretical model inspired by
observations of NGC 1333'' rather than a faithful model of the NGC
1333 cluster itself.

\begin{figure}
\figurenum{14}
{\centerline{\epsscale{0.90} \plotone{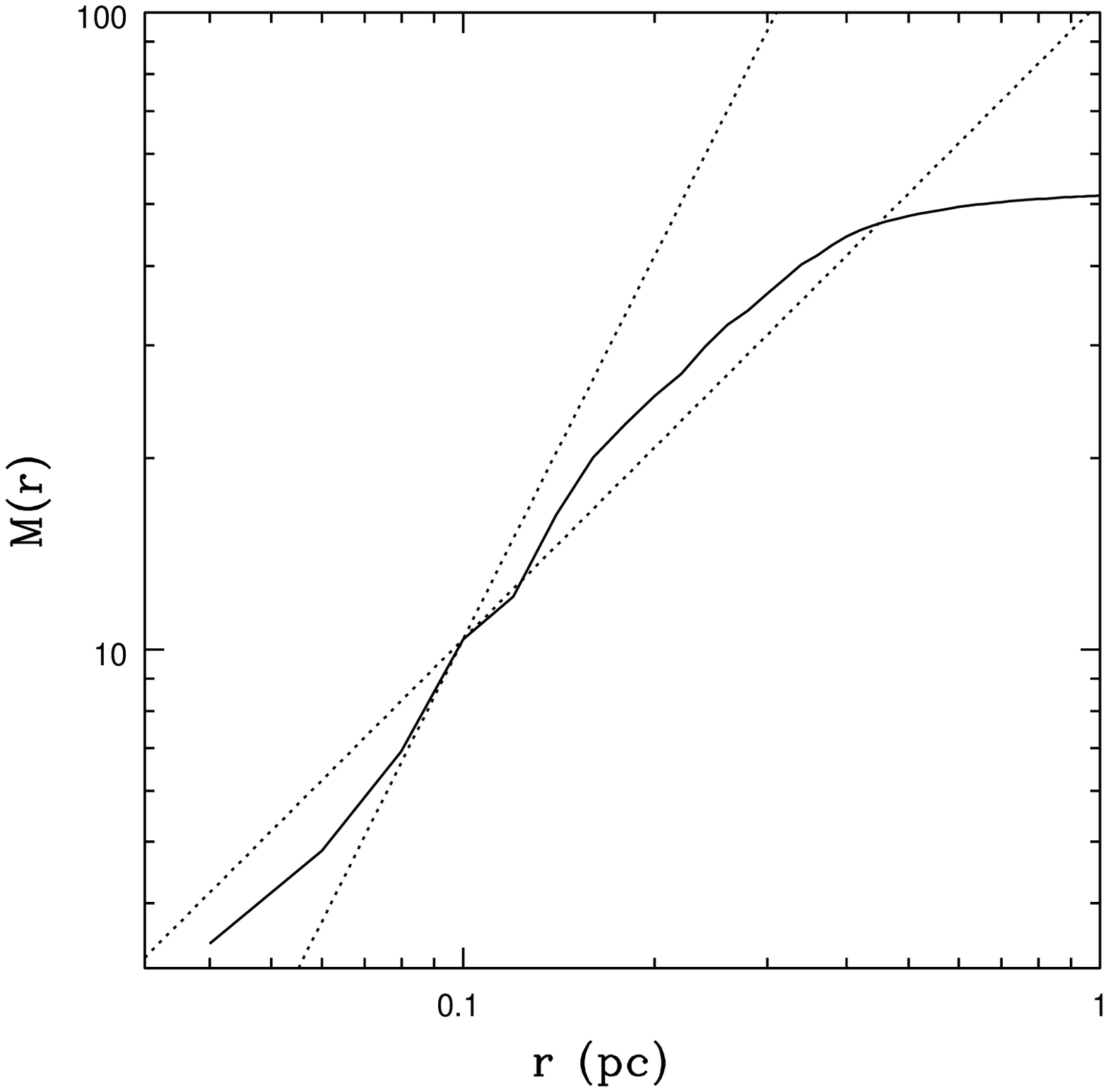} }}
\figcaption{Reconstructed radial mass profile of the young embedded cluster
NGC 1333, where $M(r)$ is given in $M_\odot$.  To obtain this profile,
the two dimensional observational map (Walsh et al. 2004ab) is converted 
into $10^5$ different realizations of the three dimensional cluster
structure (and averaged) according to the procedure outlined in the
text. The two dotted lines, included for reference, have power-law
slopes $p$ = 1 and 2, i.e., $M(r) \propto r$ (which corresponds to
$\rho(r) \propto r^{-2}$) and $M(r) \propto r^2$ (which corresponds to
$\rho(r) \propto r$).  }
\label{fig:obsdensity} 
\end{figure}

The starting conditions for the simulations are determined as follows.
For a given 2D radius $r_{2D}$ (as measured by Walsh et al. 2004ab),
we use the fact that $r_{2D} = \sin \theta r_{3D}$, and assume that
$\mu = \cos\theta$ is distributed randomly over the interval
$[-1,1]$. This procedure allows us to reconstruct the missing spatial
coordinate. The resulting radial mass profiles of the cluster are
illustrated in Figure \ref{fig:obsdensity}.  The resulting mass
profile is intermediate between that of an isothermal sphere with
$M(r) \propto r$ and a less centrally densed profile with $M(r)
\propto r^2$ (which corresponds to $\rho \sim r^{-1}$ as used in \S
2). This particular cluster is thus somewhat more centrally
concentrated than the theoretical models.  In a similar manner, we
assume that the (small) measured line of sight velocities are one
component of an isotropic (small) 3D velocity vector, and reconstruct
the missing velocity components accordingly. Since the observed speeds
are small compared to the virial speeds, the starting conditions are
much like the ``cold'' starting states of \S 2.


\begin{figure}
\figurenum{15}
{\centerline{\epsscale{0.90} \plotone{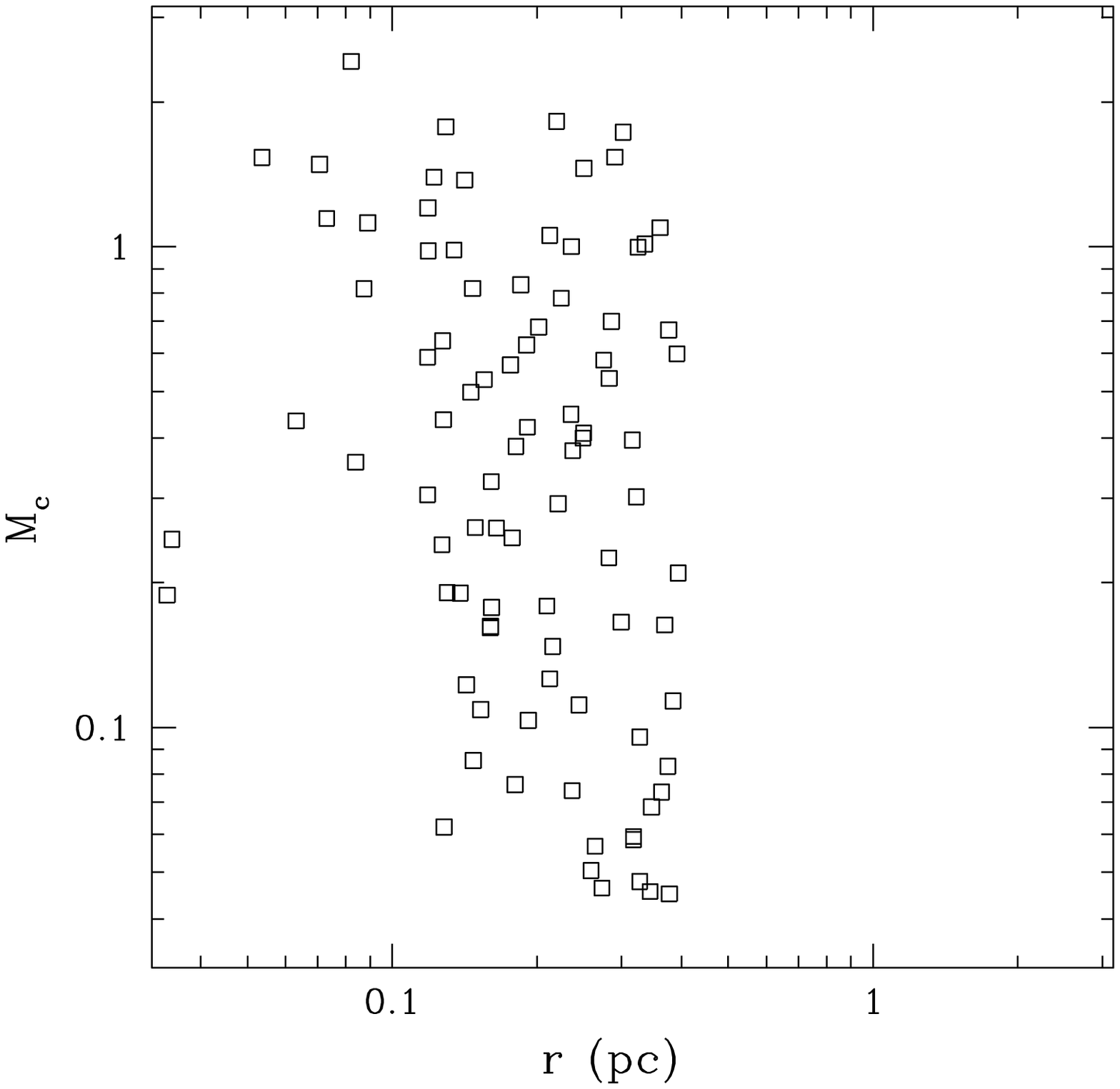} }} 
\figcaption{Mass $M_c$ of observed clumps (in $M_\odot$) as a function
of two dimensional radius for the young embedded cluster NGC 1333. The
data are taken from Walsh et al. (2004ab).  Notice that the primordial
mass segregation in this system is somewhat greater than the minimal
segregation used in the purely theoretical models (\S 2). }
\label{fig:obsmvsr} 
\end{figure}  

The observations also provide mass estimates for the clumps. For the
sake of definiteness, we assume that each clump forms a star, and that
the mass of the star is given by the mass of the clump. In actuality,
the mass of the clump is not exactly given by the mass estimated from
N$_2$H$^+$, as there is no hard boundary at the radius where the
molecule becomes too faint to be seen; this effect makes the true
clump masses larger than reported.  On the other hand, we expect some
inefficiency in the star formation process (e.g., Adams \& Fatuzzo
1996), which would make the stellar masses smaller than the clump
masses. We are thus implicitly assuming that these two effects cancel
out. The observations indicate that the clump masses in NGC 1333 are
somewhat segregated, with the more massive clumps found near the
cluster center. This trend is illustrated in Figure \ref{fig:obsmvsr}.
In this regard, the simulations of NGC 1333 differ from the purely
theoretical models of \S 2, where only minimal primordial mass
segregation was included (the most massive star was placed at the
cluster center). In addition to the mass in clumps, we include a
smooth background potential of gas analogous to the gas component used
in \S 2.

The results of the simulations for NGC 1333 are listed as the final
entries in Tables 1, 2, and 3. As expected, the output
parameters for this cluster are most like the theoretical clusters
with $N$ = 100 and cold starting conditions. However, the NGC 1333
simulations produce clusters that are somewhat more concentrated and
interactive. All of the indicators point in the same direction:
Compared to $N$ = 100 simulations with cold starting states, the NGC
1333 simulations have half mass radii that are smaller by a factor of
$\sim 1.8$, a somewhat larger fraction of stars that remain bound
(69\% versus 54\%), and a smaller isotropy parameter $\beta$ (see
Table 1). The mass profiles have roughly the same scale length $r_0
\approx 0.3 - 0.4$ pc (Table 2), and a somewhat smaller index ($p$ =
0.55 compared to 0.69), indicating a more centrally concentrated
cluster.

The largest difference between the NGC 1333 simulations and the others
is reflected in the interaction rates, where the fiducial rate
$\Gamma_0$ for NGC 1333 larger by a factor of $5-6$ (Table 3). This
higher interaction rate is a direct result of the smaller half mass
radius (a simple analytic approximation suggests that $\Gamma_0 \sim
R_{1/2}^{-7/2}$).  The characteristic interaction distance $b_C
\approx$ 238 AU, which implies that the NGC 1333 cluster facilitates
disk truncation down to radii $r_d \sim 80$ AU (still well outside the
realm of the giant planets in our solar system). Planetary analogs of
Neptune can be stripped from the smaller stars with $M_\ast$ = 0.25
$M_\odot$ and can experience large eccentricity enhancements ($e \sim
0.7$) when orbiting solar type stars (see Tables 6 -- 11).
Planetary analogs of Jupiter remain largely unperturbed around all
stars. This level of disruption is somewhat higher than found earlier
for the purely theoretical clusters, but still remains modest.

One important lesson resulting from this case study of NGC 1333 is the
extent to which initial conditions can affect forming planetary systems.  
Compared to the starting conditions used for the simulations in \S 2
(where these starting conditions were motivated by observational
surveys of cluster conditions, e.g., Figs. \ref{fig:probn} and
\ref{fig:rvsn}), the NGC 1333 starts with a higher degree of central
concentration and a greater amount of primordial mass segregation. The
result is a more compact cluster (smaller $R_{1/2}$) and hence a
higher interaction rate. In addition, the cold starting condition
allows stars to fall inside much of the original gas and this geometry
enhances cluster survival after gas removal (compare Adams 2000 with
Geyer \& Burkert 2001).  In order to fully determine the effects of
the cluster environment on forming solar systems, we need to determine
the range of starting density profiles and mass segregation.

\section{Conclusion} 

This paper has explored the early dynamical evolution of embedded
stellar groups and clusters with stellar membership in the range $N$ =
100 -- 1000. This work includes $N$-body simulations of the dynamics,
compilations of the distributions of FUV luminosities and fluxes, the
calculation of scattering cross sections for young planetary systems,
and an application to the observed embedded cluster NGC 1333. Our main
conclusion is that clusters (with the range of properties considered
here) have relatively modest effects on star and planet formation.
The interaction rates and radiation levels are low, so that forming
stars and their accompanying planetary systems are largely unperturbed
by their environment. This finding, in turn, implies that cluster
structure is due primarily to the initial conditions, rather than
interactions.  These results can be summarized in greater detail as
follows:

[1] In order to obtain good statistics for our output measures, we
have performed 100 realizations of each set of initial conditions for
groups/clusters with $N$ = 100, 300, and 1000. In addition to
considering different cluster sizes $N$, we consider both ``virial'' and
``cold'' initial conditions. These simulations show a significant
difference between the two types of starting conditions. Compared with
``virial'' initial conditions (near virial equilibrium), ``cold''
clusters are more centrally concentrated, retain more of their stars
for longer times, and exhibit more radial velocity distributions
(Tables 1 and 4).  As expected, all clusters lose stars and
gradually spread out with time.  This behavior is quantified by
finding the average time evolution of each group/cluster type using
100 realizations of each set of initial conditions (Fig.
\ref{fig:output} and Table 1).  We also provide quantitative
descriptions of these systems by finding the mass profiles of the
clusters (Fig.  \ref{fig:radial}, eq.  [\ref{eq:mfit}], and Table 2)
and the distributions of close encounters (Fig. \ref{fig:interact},
eq.  [\ref{eq:ratenumer}], and Table 3). All of these quantities can
be used in a variety of other contexts to test further the effects of
the cluster environment on the processes of star and planet formation.

[2] We have calculated the FUV radiation expected from this class of
groups and clusters. This issue involves (at least) three separate
distributions: Clusters of a given size $N$ display a wide
distribution of FUV luminosities due to incomplete sampling of the
stellar IMF; we have determined this distribution $P(L_{FUV})$ as a
function of $N$ (Figs. \ref{fig:lfuv} and \ref{fig:lfuvdist}).
Clusters themselves come in a distribution of sizes $P(N)$
(Fig. \ref{fig:probn}) and we have found the distribution of FUV
luminosities sampled over all clusters (Fig. \ref{fig:ldistalln})
using the observed range of cluster radii $R$ (Fig. \ref{fig:rvsn}).
Finally, the stars within a cluster explore a range of radial
positions, which in turn specify the distribution of radial positions
$P(r)$ in the cluster.  These three probability distributions [$P(N)$,
$P(L_{FUV})$, $P(r)$] jointly determine the composite distribution of
FUV fluxes that impinge upon the composite ensemble of forming solar
systems (shown in Fig.  \ref{fig:fluxdist}).  The median FUV flux for
the composite distribution is only $G_0 \approx 900$, which is not
intense enough to evaporate disks orbiting solar-type stars (over 10
Myr) for the range of radii of interest for planet formation ($r \le
30$ AU). We have also found the fluxes averaged over individual orbits
within the clusters as a function of orbital energy and angular
momentum (eqs. [\ref{eq:meanflux} -- \ref{eq:meanflux2}]). The results
of this section imply that FUV radiation in clusters does not
generally inhibit planet formation. In addition, the distributions
found here can be used to determine the radiation exposure for forming
solar systems in a variety of other contexts.

[3] We have calculated the cross sections for the interaction of newly
formed solar systems with passing binaries (Tables 6 -- 11) using
an ensemble of $\sim10^5$ Monte Carlo scattering experiments.  These
cross sections, in conjunction with the interaction rates determined
via the $N$-body simulations, show that the typical solar system is
not greatly affected by scattering interactions within its birth
aggregate. The ``typical'' star within a cluster of size $N$ = 100 --
1000 will experience approximately one close encounter within a
distance $b_C$ over a 10 Myr window of time. We find that $b_C$ = 700
-- 4000 AU for the systems considered here.  This passage is not close
enough to appreciably enhance the eccentricity of Neptune in our solar
system. The mildest disruption event considered here is the increase
in eccentricity of a Neptune-analog planet orbiting a 0.125 $M_\odot$
star; the cross section for this event is $\cross \sim 2 \times 10^5$
AU$^2$ (Table 10), requiring a closest approach distance of $\sim250$
AU.  Similarly, disks are truncated by passing stars down to radii of
$\sim1/3$ of the closest approach distance (Kobayashi \& Ida 2001), so
the disks in these clusters are expected to be limited to 230 -- 1300
AU, much larger than the regimes of interest for planet formation.
Our main conclusion is that planet forming disks and newly formed
solar systems generally survive their birth aggregates with little
disruption. In addition, the cross sections calculated herein can be
used to study solar system disruption in a wider range of contexts and
environments. For example, planet formation can potentially be induced
by weak scattering encounters (Thies et al. 2005).  As another
application, we note that some star forming regions are reported to
have higher binary fractions than the field.  As a result, one issue
is whether or not the cluster environment can disrupt binaries (e.g.,
Kroupa et al. 2003 and references therein). Our results imply that the
clusters considered here do not facilitate the disruption of binaries,
except for those that begin with separations greater than $\sim$1000
AU. If the primordial period distribution is similar to that measured
in the field, only $\sim1/7$ of binaries would be affected by
scattering interactions (of course, more binary disruption would occur 
if these clusters did not suffer an early demise due to gas expulsion 
at $t$ = 5 Myr). 

[4] We have performed an ensemble of 100 simulations of the observed
young embedded cluster NGC 1333, where we start with observed
positions in the plane of the sky and line of sight velocity
components, and then reconstruct the remaining phase space
coordinates. This set of simulations is used to construct the output
measures for clusters of this type and we use the results to assess
the impact of the background environment on star and planetary sytems
forming within this type of group/cluster. This cluster is most like
the $N$ = 100 cold simulations performed in \S 2. Compared to the
purely theoretical simulations, NGC 133 has more primordial mass
segregation and a smaller half-mass radius $R_{1/2}$. This property
leads to a somewhat larger bound fraction $f_b$ and a higher
interaction rate $\Gamma$ compared to the $N$ = 100 simulations with
cold starts. Nonetheless, the overall amount of disruption is small
(e.g., circumstellar disks are truncated down to $\sim80$ AU, well
outside the region where giant planets form) so that the cluster
environment has only a modest effect on star and planet formation.

This paper represents an assessment of dynamical effects in six
classes of young embedded clusters. The treatment is comprehensive in
that we run 100 $N$-body simulations for each type of cluster in order
to obtain robust statistical descriptions, and we assess the effects
of FUV radiation and solar system scattering on forming solar systems.
On the other hand, the parameter space available to such clusters is
enormous and a great deal of additional work remains to be done. For
example, the simulations in this study were started with cluster sizes
$\rstar$ near the low end of the observed range (the lower dashed
curve in Fig. \ref{fig:rvsn}) and gas removal times (5 Myr) near the
high end of the observed range (Lada \& Lada 2003). These choices tend
to make the clusters denser and long-lived, which makes the effects of
interactions and radiation more important. Since we find that
interactions and radiation have only modest effects on planet forming
disks, we can consider this conclusion as conservative.  However, a
more detailed treatment of gas removal, including shorter lifetimes
and more realistic time dependence (not a step function in time), is
warranted. 

A number of additional processes may also affect cluster evolution and
should be studied; these include the role played by additional
primordial mass segregation (beyond the minimal treatment used here),
non-spherical starting conditions for both the stars and gas, and the
effects of primordial binaries. Mass segregation -- both primordial
and evolutionary -- may be particularly interesting, as suggested by
our simulations motivated by NGC 1333.  Our work to date indicates
that the disruption of planetary systems is a sensitive function of
the mass $M_\ast$ of the central star (e.g., scattering cross sections
scale approximately as $\cross \sim M_\ast^{-1/2}$ and FUV radiation
truncates disks approximately at $r_d \sim M_\ast$) and a sensitive
function of location within the cluster (both the FUV flux and
interaction rates are much greater near the cluster center). If
substantial mass segregation is present during the $\sim$10 Myr while
the clusters remain intact, the larger stars will be closer to the
center where they are exposed to greater probability of disruption,
and the smaller stars will be farther out and relatively safer. The
degree to which this effect occurs should be quantified in future
studies.

Finally, this work emphasizes the fact that cluster environments
display a distribution of properties and the full distributions must
be considered in order to assess their effects on forming stars and
planets. Some previous studies (e.g., Bonnell \& Bate 2002) have
focused on the densest regions of large clusters where the interaction
rates are high and the background environment has an important effect
on star formation. Although most clusters have a central zone of high
interaction, for the clusters considered here most stars do not live
in the highly interactive zone. It is thus crucial to determine the
full distribution of environmental properties that forming stars are
exposed to, including how often the various environments arise.
Clusters are sampled from a distribution of sizes $P(N)$.  For a given
size $N$, clusters have a range of radial sizes $P(\rstar)$, a range
of starting speeds and hence a distribution of virial parameters
$P(Q)$, and display a distribution of FUV luminosities $P(L_{FUV})$.
For given sizes $\rstar$ and $N$, and a given starting condition $Q$,
stellar members explore a distribution of radial positions $P(r)$
within the cluster. The methods developed in this paper show that we
can find the distributions of luminosities (Fig. \ref{fig:ldistalln}),
radial positions (analogous to Fig. \ref{fig:radial}), closest
approaches (Fig.  \ref{fig:interact}), and other quantities of
interest from a given set of starting conditions. Perhaps the most
important goal of future studies is thus to make a better
observational determination of the distributions $P(N)$, $P(Q)$, and
$P(\rstar)$, which would allow for a more complete assessment of the
effects of the cluster environment on star and planet formation.
 
\vskip 0.35truein 

\centerline{\bf Acknowledgment} 

We would like Lori Allen, Tony Bloch, Gus Evrard, and Tom Megeath for
many beneficial discussions. This work was supported at the University
of Michigan by the Michigan Center for Theoretical Physics, and by
NASA through the Terrestrial Planet Finder Mission (NNG04G190G) and
the Astrophysics Theory Program (NNG04GK56G0). This work was supported
at Xavier University through the Hauck Foundation. Finally, P. Myers
acknowledges Grant NAG5-13050 from the NASA Origins of Solar Systems
Program. 

\newpage

\end{document}